\documentclass[a4paper,onecolumn,11pt]{IEEEtran}

\include{preamble}

\usepackage{epstopdf}

\begin{document}

\title{Probability Estimates for Fading and Wiretap Channels from Ideal Class Zeta Functions}

\author{David Karpuk\thanks{D. Karpuk and C. Hollanti  are with the Department of Mathematics and Systems Analysis, P.O. Box 11100, FI-00076 Aalto University, Finland  (e-mails: camilla.hollanti@aalto.fi, david.karpuk@aalto.fi).}, Anne-Maria Ernvall-Hyt\"onen,\thanks{A.-M. Ernvall-Hyt\"onen is with the Department of Mathematics and Statistics, FI-00014 University of Helsinki, Finland (e-mail: anne-maria.ernvall-hytonen@helsinki.fi).}  Camilla Hollanti, and Emanuele Viterbo\thanks{E. Viterbo is with  the Department of Electrical and Computer Systems Engineering, PO Box 35, Monash University, Clayton, Victoria 3800, Australia (e-mail: emanuele.viterbo@monash.edu).}\thanks{The research of D. Karpuk is supported by Academy of Finland grant \#268364 and the Magnus Ehrnrooth Foundation, Finland.  C. Hollanti is supported by the Academy of Finland grants \#276031, \#282938, and \#283262, and by Magnus Ehrnrooth Foundation, Finland. 
A.-M. Ernvall-Hyt\"onen is supported by the Academy of Finland grants \#138337 and \#138522.}\thanks{Part of this work was performed at the Monash Software Defined Telecommunications Lab
and was supported by the Monash Professional Fellowship and the Australian Research
Council under Discovery grants ARC DP 130100103. This research was partly carried while  C. Hollanti was visiting E. Viterbo at the Monash University in 2011.}\thanks{The support from the European Science Foundation under the ESF COST Action IC1104 is
also gratefully acknowledged.}\thanks{Part of the results  in Section \ref{bounds1} were presented at ICUMT 2011 \cite{ICUMT11}.}\thanks{AMS Classifications 14G50, 14G25.}
}
\maketitle

\begin{abstract}
In this paper, new probability estimates are derived for ideal lattice codes from totally real number fields using ideal class Dedekind zeta functions. In contrast to previous work on the subject, it is not  assumed that the ideal in question is principal.  In particular, it is shown that the corresponding inverse norm sum depends not only on the regulator and discriminant of the number field, but also on the values of the ideal class Dedekind zeta functions.  Along the way, we derive an estimate of the number of elements in a given ideal with a certain algebraic norm within a finite hypercube.  We provide several examples which measure the accuracy and predictive ability of our theorems.

\end{abstract}


\IEEEpeerreviewmaketitle

\section{Introduction}

It has been well-known for many years that number field lattice codes provide an efficient and robust means for many applications in wireless communications. We refer to \cite{OV} for a thorough introduction to the topic. More recently,  number field based codes have been studied in conjunction with fading wiretap channels. Gaussian and fading wiretap channels have been considered in \cite{Hell_Gaussianwire,belfisoleoggis,belfisole,belfioggiswire,oggis_new}.  In \cite{BO_wiretap} the authors  propose using lattice codes constructed from totally real number fields, which also form the basis for our study and constructions. The behavior of the probability of Eve's correct decision depends on the \emph{inverse norm sum}, which is our principal object of study\footnote{It was  also pointed out in \cite{oggis_new} that the approximation of Eve's probability  by the inverse norm sum can be sometimes quite loose. This is a general feature of the well-known union bound technique, also used here to bound the probability.  Nevertheless, the inverse norm sum enables clean algebraic analysis and comparison of different lattices without having to start with heavy simulations and, at least in an appropriate SNR range, helps  to predict the performance order of different codes, if not the actual performance. In particular, it does enable us to pick the best code when the union bound is used as a design criterion.}.

  The inverse norm sum has been analyzed in some example cases in \cite{ITW_camiame}.  This paper can be seen, on one hand, as a continuation of \cite{ITW_camiame,ICUMT11}, where analysis on lattice codes in fast and block fading channels was carried out based on various explicit code constructions and, on the other hand, a generalization of the number field case of \cite{ISIT11-roopefransu,ITW11-roopefransu}, where Vehkalahti \emph{et al.} showed how the unit group and diversity-multiplexing gain trade-off (DMT) of division algebra-based space-time codes are linked to each other through inverse determinant sums, and also demonstrated the connection to zeta functions and point counting.

Our work differs from this and the subsequent work \cite{roopelaura_ISIT, roopefransujournal} in that we consider non-principal ideals and provide a more precise expression for the inverse norm sum.  Our results allow analysis of both the pairwise error probability of the Rayleigh fading channel as well as the probability of an eavesdropper's correction decision in a wiretap channel.   While in \cite{roopefransujournal} the authors concentrate on the number of units in a finite spherical subset of a lattice, here we estimate each individual term in the inverse norm sum by estimating the number of points of a given norm in a cubic constellation. The main conclusion of our approach is that the inverse norm sum is determined by both the density of the units (i.e.\ the regulator) and values of the ideal class Dedekind zeta functions.  These zeta values can vary wildly between ideal classes and even between ideals of the same norm; see the examples following Theorem \ref{main_theorem}.  The dependence on the zeta values is important for non-principal ideals and principal ideals in fields with class number larger than $1$.

Our main theorem, Theorem \ref{main_theorem}, can be summarized as follows.  Let $K/\Q$ be a totally real number field of degree $n$, and let $\frak{a}\subseteq\mathcal{O}_K$ be an ideal.  Let $\Lambda = (\frak{a},q_\alpha)$ be an ideal lattice, with twisted canonical embedding $\psi_\alpha:\frak{a}\rightarrow \R^n$, and scaled by a constant $\kappa$ so that $\vol(\Lambda) = 1$.  Define the inverse norm sum
\begin{equation}\label{first}
S(\Lambda,s,R) = \sum_{\substack{0\neq x\in\Lambda \\ ||x||_{\infty}\leq R}} \prod_{i = 1}^n\frac{1}{|x_i|^s} 
= \frac{1}{k^{ns}|N(\alpha)|^{s/2}}\sum_{\substack{0\neq x\in\frak{a} \\ ||\psi_\alpha(x)||_{\infty}\leq R/\kappa}} \frac{1}{|N(x)|^s} 
\end{equation}
where $N:K\rightarrow \Q$ is the field norm.  Then
\begin{equation}\label{second}
\boxed{S(\Lambda,s,R) = \frac{w_K|D_K|^{s/2}}{R_K}\zeta_K^{[\frak{a}]^{-1}}(s)c_n\log(R)^{n-1} + O(\log(R)^{n-2})}
\end{equation}
where $c_n$ is a constant depending only on $n$, $[\frak{a}]$ denotes the class of $\frak{a}$ in the ideal class group of $K$, and $\zeta_K^{[\frak{a}]^{-1}}(s)$ is the ideal class Dedekind zeta function associated with the inverse class $[\frak{a}]^{-1}$ (cf.\ \eqref{ideal_zeta}).  The other constants are standard number-theoretic invariants of $K$, defined in the next section.  We do \emph{not} assume $\frak{a}$ is a principal ideal as is often done in the literature, and thus one cannot reduce to the case $\Lambda = (\mathcal{O}_K,q_\alpha)$ as is often done.  The choice of the norm $||\cdot||_\infty$, i.e.\ cubic shaping, is mostly a convenience which simplifies our proof of Theorem \ref{bkR_estimate}. Cubic shaping is also often preferred in practice as it simplifies bit labeling.  It is easy to see that our results apply to any norm $||\cdot||_p$, i.e.\ for example to spherical shaping as well. 

From an engineering perspective, normalizing the volume of $\Lambda$ so that $\vol(\Lambda) = 1$ is necessary to compare inverse norm sums between lattices of the same dimension.  This is somewhat of a cosmetic alteration mathematically, but it does help tease out the exact invariants of $K$ and $[\frak{a}]$ on which $S(\Lambda,s,R)$ depends.  Pulling off the coefficient of $\log(R)^{n-1}$ in our expression for $S(\Lambda,s,R)$ (and dividing by $c_n$) allows us to define the following invariant, which predicts the growth of $S(\Lambda,s,R)$ as a function of $R$:
\begin{equation}
\boxed{\sigma(K,[\frak{a}],s) = \frac{w_K|D_K|^{s/2}}{R_K}\zeta_K^{[\frak{a}]^{-1}}(s)}
\end{equation}
If an ideal lattice defined by a principal ideal $\frak{a} = (\alpha)$ is normalized so that $\vol(\Lambda) = 1$, the design criterion given by the minimum product distance reduces to $d_{p,\min}(\Lambda) = |D_K|^{-1/2}$ (see \cite[Theorem 6.1]{OV}).  Thus finding a number field $K$ and an ideal class $[\frak{a}]$ which minimizes $\sigma(K,[\frak{a}],s)$ is a subtler task.  We study how this invariant varies with $K$ and $[\frak{a}]$ in the examples following our Theorem \ref{main_theorem}.  We do not assume $\Lambda$ to be cubic, and thus if one wants to work with rotated versions of $\Z^n$ as in \cite{OV} one must still find appropriate $\frak{a}$ and $\alpha$.

In general the estimation error in our Theorem \ref{bkR_estimate} and Theorem \ref{main_theorem} increases with the dimension of the lattice. Notice that the lattice dimension is not limiting the data rate as we can always increase the constellation size by choosing a bigger hypercube, which decreases the relative estimation error since the edge error effect becomes more negligible. Another limitation to the lattice dimension is forced by decoding, since the complexity of any maximum-likelihood (ML) decoder such as a sphere decoder grows exponentially with the lattice dimension.  

We would like to mention previous work which fits nicely into the theoretical framework of our paper.  We show experimentally that for the unimodular lattices from quadratic fields and quartic fields studied in \cite{ducoat_oggier}, the coefficient $\sigma(K,[\frak{a}],s)$ predicts the relative sizes of the inverse norm sums.   This gives a broader theoretical foundation to the work contained in \cite{ducoat_oggier}, as well as explains the heavy dependence of the inverse norm sum on the discriminant mentioned therein.  The authors of \cite{ong_oggier} explore real cyclotomic number fields with few elements of small norm, to attempt to minimize the corresponding inverse norm sum.  In the context of our results, this is equivalent to minimizing the zeta value $\zeta_K^{[1]}(s) = \sum a^{[1]}_k/k^s$, where $a^{[1]}_k$ is the number of principal ideals of norm $k$.  In terms of pure number theory, an estimate of the number of units under the canonical embedding in a box of fixed size has been given in \cite{EV,EVLO}.  As part of the proof of our main theorem, we have given in Theorem \ref{prinideal_coeffs} similar estimates to the number of lattice points of given norm contained in a given ideal under the canonical embedding.

 The organization and main contributions of the rest of the paper are as follows:
\begin{itemize} 
\item The next two sections are devoted to the necessary number theoretic and wireless communications background. 
\item In Section \ref{bounds1} we derive elementary bounds on the inverse norm sums of ideal lattices.  For the sake of simplicity, we use the unnormalized, untwisted canonical embedding of an ideal in this section.
\item In Section \ref{geometric} we derive an estimate of the number of elements $x$ in the (unnormalized, untwisted) ideal lattice of norm $k$ and $||x||_{\infty}\leq R$. We provide examples demonstrating the accuracy of this estimate, showing that the estimate is very good when the dimension is relatively low and hence the decoding delay is short.
\item Section \ref{main_section} is devoted to proving our main theorem, Theorem \ref{main_theorem}, by using the results of the previous section.  We show by example that our theorem predicts the relative behavior of the inverse norm sums well.  We use our main theorem to demonstrate how the growth of inverse norm sums of non-principal ideal lattices varies with the ideal class, and provide examples.
\item We use the appendix to prove a technical lemma which bounds the tail of the ideal class Dedekind zeta function, thus also gives a bound to the error term in our estimate.  
\item We provide conclusions in the final section, which discuss potential generalizations to fractional ideals and to CM-fields, as well as further future work.
\end{itemize}

\section{Algebraic preliminaries}
\label{preli}

In this section we review the essential number theoretic concepts.  As a catch-all reference for algebraic number theory, we recommend \cite{lang}.
\subsection{Number Field Basics}
A \emph{number field} $K$ is a finite extension of $\Q$.  The \emph{ring of integers} $\mathcal{O}_K$ of $K$ is the integral closure of the ring $\Z$ in $K$, and it is a $\Z$-module of rank equal to $n = [K:\Q]$.  A \emph{real embedding} of $K$ is a field homomorphism $\sigma:K\hookrightarrow \R$, and a \emph{complex embedding} is a field homomorphism $\sigma:K\hookrightarrow \C$ such that $\sigma(K)\not\subseteq \R$.  A number field is \emph{totally real} if it admits no complex embeddings.  If $r_1$ (resp.\ $r_2$) denotes the number of real (resp. complex) embeddings, then $r_1 + 2r_2 = n$, so that $r_1 = n$ if $K$ is totally real.

Lattices will play a key role throughout the paper, so let us recall the notion of a lattice.  For any $n> 0$, a \emph{lattice $\Lambda$ of rank $t\leq n$} is a discrete subgroup of the real vector space $\R^n$, such that $\R\otimes_\Z\Lambda\cong \R^t$.  Equivalently, $\Lambda$ is the $\Z$-span of $t$ vectors in $\R^n$ which are linearly independent over $\R$.  The number $t$ is the \emph{rank} of the lattice, and if $t = n$ we say that $\Lambda$ is \emph{full rank}.  If a full-rank lattice is the $\Z$-span of the column vectors $v_1,\ldots,v_n$, then we define $\vol(\Lambda) = |\det[v_1,\ldots,v_n]|$, which can be shown to be independent of the choice of $v_i$.

Let $K/\Q$ be a number field of degree $n$,  $\sigma_1,\ldots, \sigma_{r_1}$ its real embeddings, and $\sigma_{r_1+1},\ldots,\sigma_{r_1 + r_2}$ and set of representatives of the complex embeddings modulo complex conjugation. The \emph{canonical embedding} $\psi:K\hookrightarrow \R^{r_1}\times \C^{r_2}$ is defined by the map
\begin{equation}\label{can-emb}
\psi(x)=(\sigma_1(x),\ldots,\sigma_{r_1}(x),\sigma_{r_1+1}(x),\ldots,\sigma_{r_1 + r_2}(x))\in \R^{r_1}\times \C^{r_2},
\end{equation}
One can show that $\psi(\frak{a})$ is a full-rank lattice in $\R^{r_1}\times \C^{r_2} = \R^{r_1 + 2r_2} = \R^n$, for any ideal $\frak{a}\subseteq\mathcal{O}_K$.  If $\omega_1,\ldots,\omega_n$ is a $\Z$-basis of $\mathcal{O}_K$, then the \emph{discriminant} $D_K$ is defined by $D_K = \det((\sigma_i(\omega_j))_{1\leq i,j\leq n})^2$, so that $|D_K| = \vol(\psi(\mathcal{O}_K))^2$.

If $\sigma_1,\ldots,\sigma_n$ denote all embeddings of $K$ into $\C$, then we define the \emph{norm map} $N:K\rightarrow \Q$ by
\begin{equation}
N(x)=\prod_{i=1}^n\sigma_i(x).
\end{equation}
Thus if $K/\Q$ is totally real, we have $N(x) = \prod_{i = 1}^n\psi(x)_i$.  If $\frak{a}\subseteq \mathcal{O}_K$ is an ideal, then we define
\begin{equation}
N(\frak{a}) = \#(\mathcal{O}_K/\frak{a})
\end{equation}
to be the cardinality of the corresponding quotient ring.  When $\frak{a} = (\alpha)$ is a principal ideal, one can check that $|N(\alpha)| = N((\alpha))$ and thus the two definitions coincide.  The norm is multiplicative in the sense that if $\frak{a}$ and $\frak{b}$ are two ideals of $\mathcal{O}_K$, then $N(\frak{a}\frak{b}) = N(\frak{a})N(\frak{b})$.

\begin{theorem}\emph{(Dirichlet Unit Theorem, \cite[Chapter V \S1]{lang})}\label{unitthm}
Let $K$ be a number field and let $r = r_1 + r_2 -1$.  Then there are units $\epsilon_1,\ldots,\epsilon_{r}\in\OO_K^\times$ such that
\begin{equation}
\OO_K^\times \cong \mu_K\times \langle\epsilon_1\rangle\times\cdots\times \langle\epsilon_{r}\rangle
\cong \mu_K\times \Z^{r},
\end{equation}
where $\mu_K$ is the group of roots of unity in $K$. The $\epsilon_j$ are called a \emph{fundamental system of units} for $K$.
\end{theorem}

Let $\{\epsilon_1,\ldots,\epsilon_{r}\}$ be a fundamental system of units for $K$.  If $|\cdot|$ denotes the usual absolute value on $\C$, consider the matrix
\begin{equation}
A=(\log|\sigma_j(\epsilon_i)|_j)
\end{equation}
for $1\leq i\leq r$ and $1\leq j\leq r_1+r_2$, where we have used the notation
\begin{equation}
|x|_j=\left\{\begin{array}{ll}
|x| & \textrm{ if } 1\leq j\leq r_1,\\
|x|^2 & \textrm{ if } r_1 + 1\leq j\leq r_1+r_2.
\end{array}\right.
\end{equation}
The \emph{regulator} $R_K$ is the absolute value of the determinant of any $r\times r$ minor of $A$. It is independent of the choice of the fundamental system of units and the choice of minor.  The volume of the fundamental parallelotope of the log-lattice $\Lambda_{\log}$ generated by $A$  is expressed in terms of the regulator as 
\begin{equation}
\vol(\Lambda_{\log})=R_K\sqrt{r_1+r_2}
\end{equation}
In the case of a totally real number field we have $\vol(\Lambda_{\log})=R_K\sqrt{n}$. The regulator is a positive real number that in essence is inversely proportional to the density of the units, and can easily be computed using Sage  \cite{sage} when the dimension is not too big.

\subsection{Ideal Lattices}

The lattice codes we use are constructed as follows.  Let $K/\Q$ be a totally real number field of degree $n$.  An \emph{ideal lattice} $\Lambda = (\frak{a},q_\alpha)$ consists of the following data: an ideal $\frak{a}\subseteq\mathcal{O}_K$, and a trace form
\begin{equation}
q_\alpha:\frak{a}\times\frak{a}\rightarrow \Z,\quad q_\alpha(x,y) = \text{Tr}(\alpha xy), \text{ for $x,y\in\frak{a}$}
\end{equation}
where the \emph{twisting element} $\alpha\in K$ is totally positive, in the sense that $\sigma_i(\alpha)\in\R_{>0}$ for all embeddings $\sigma_i:K\hookrightarrow \R$.  Given the data of an ideal lattice $\Lambda = (\frak{a},q_\alpha)$, the actual lattice in question is defined by the \emph{twisted canonical embedding} $\psi_{\alpha}$, given by
\begin{equation}
\Lambda = \psi_{\alpha}(\frak{a}) = \psi(\frak{a})\cdot \diag\left(\sqrt{\sigma_1(\alpha)},\ldots,\sqrt{\sigma_{n}(\alpha)}\right)
\end{equation}
where $\psi:K\hookrightarrow\R^n$ denotes the canonical embedding.  More explicitly, if $x\in\frak{a}$, the corresponding lattice vector in $\R^n$ is given by
\begin{equation}\label{embvector}
\psi_\alpha(x) = \left(\sqrt{\sigma_1(\alpha)}\sigma_1(x),\ldots,\sqrt{\sigma_n(\alpha)}\sigma_n(x)\right)
\end{equation}
In what follows we will use the fact that $\prod_{i = 1}^n |\psi_\alpha(x)_i| = |N(\alpha)|^{1/2}|N(x)|$.

\subsection{The Class Group and Ideal Class Dedekind Zeta Functions}

A \emph{fractional ideal} $\frak{a}$ of $K$ is an $\mathcal{O}_K$-submodule of $K$ such that there exists $x\in \mathcal{O}_K$ with $x\frak{a}\subseteq\mathcal{O}_K$.  The group of non-zero fractional ideals forms an abelian group $I_K$ under multiplication, and the principal fractional ideals $P_K$ form a subgroup.  The quotient $C_K := I_K/P_K$ is the \emph{class group} of $K$, and it is known to be finite.  If $\frak{a}$ is a fractional ideal of $K$ (e.g.\ an ideal of $\mathcal{O}_K$) we denote by $[\frak{a}]$ its class in $C_K$.  The \emph{class number} $h_K$ of $K$ is the cardinality of the group $C_K$.  The class number measures, in some sense, the failure of the ring $\mathcal{O}_K$ to be a PID.

\begin{definition}(\emph{Ideal class Dedekind zeta function, \cite[Chapter VIII \S2]{lang}}) \label{zeta}
Let $[\frak{a}]\in C_K$ be an ideal class in $K$.  The \emph{ideal class Dedekind zeta function} of $[\frak{a}]$, and the \emph{Dedekind zeta function of $K$}, are defined respectively by
\begin{equation}\label{ideal_zeta}
\zeta_K^{[\frak{a}]}(s)=\sum_{\substack{\frak{b}\subseteq\OO_K \\ [\frak{b}] = [\frak{a}]}}\frac1{N(\frak{b})^s} = \sum_{k= 1}^{\infty}\frac {a^{[\frak{a}]}_k}{k^s},\quad\text{and}\quad \zeta_K(s) = \sum_{[\frak{a}]\in C_K}\zeta_K^{[\frak{a}]}(s)
\end{equation}
where $a^{[\frak{a}]}_k$ is the number of integral ideals of norm $k$ in the same class as $\frak{a}$ in $C_K$.
\end{definition}  

We refer to the coefficients $a^{[\frak{a}]}_k$ as \emph{Dirichlet coefficients}.  It is well-known that $\zeta^{[\frak{a}]}_K(s)$ converges for $\Re(s)>1$.   For the applications under study the interesting values are $s=2$ (the pairwise error probability) and $s = 3$ (the eavesdropper's error probability).  If $\mathcal{O}_K$ is a PID then there is only one ideal class and $\zeta_K^{[1]}(s) = \zeta_K(s)$.  In term of the applications we consider, working with $\zeta_K^{[\frak{a}]}(s)$ instead of $\zeta_K(s)$ is necessary if one wants to consider ideal lattices defined by non-principal ideals, or even principal ideals in number fields $K$ with $h_K > 1$.  Numerically evaluating the ideal class zeta functions can be done easily in Sage \cite{sage}.

We mention the following theorem to demonstrate how the above invariants of $K$ are all related to each other.  The resemblance of the coefficient of $\log(R)^{n-1}$ in our Theorem \ref{main_theorem} to the residues of the ideal class zeta functions is also suggestive of a potential deeper connection between the inverse norm sums and the Class Number Formula.
\begin{theorem}\emph{(Class Number Formula, \cite[Chapter VIII \S2, Theorem 5]{lang})}
Let $K$ be a number field with $r_1$ real embeddings, $2r_2$ complex embeddings, discriminant $D_K$, regulator $R_K$, class number $h_K$, and let $w_K$ be the number of roots of unity in $K$.  Then $\zeta_K^{[\frak{a}]}(s)$ has a simple pole at $s=1$, with residue
\begin{equation}\label{ideal_cnf}
\text{Res}_{s=1}\zeta_K^{[\frak{a}]}(s) = \frac{2^{r_1}(2\pi)^{r_2}R_K}{w_K\sqrt{|D_K|}}\quad\text{so that}\quad \text{Res}_{s = 1}\zeta_K(s) = \sum_{[\frak{a}]} \text{Res}_{s=1}\zeta_K^{[\frak{a}]}(s) = \frac{2^{r_1}(2\pi)^{r_2}h_KR_K}{w_K\sqrt{|D_K|}}.
\end{equation}
\end{theorem}

\section{Probability expressions and inverse norm sums}
\label{probinv}

Our main references for the wireless communications background are \cite{OV}, which introduces ideal lattices in the context of lattice coding, and \cite{BO_wiretap}, which shows that the inverse norm sum determines the probability of an eavesdropper's correct decision in a wiretap channel.

\subsection{The Rayleigh fading channel}

Following \cite{OV}, we define a Rayleigh fading channel by the channel equation
\begin{equation}
y = hx + z
\end{equation}
where $x\in \R^n$ is the vector intended for transmission, $h = \diag(h_i)$ is a fading diagonal matrix with $h_i$ a Rayleigh random variable with $\mathbf{E}(h_i^2) = 1$, $z = (z_i)$ is additive white Gaussian noise with $z_i = N(0,\sigma^2)$, and $y$ is the received signal.

The vector $x$ is selected from a finite constellation $\mathcal{C}\subset\R^n$, which in our case will be a subset of a lattice $\Lambda$ of the form $\{ x\in \Lambda\ |\ ||x||\leq R\}$ for some $R> 0$ and some norm $||\cdot ||$.  One common judge for performance is the \emph{pairwise error probability}, or PEP, denoted by $P_e$ and which measures the probability that the received signal $y$ is decoded as some $x'\neq x$ instead of the intended $x$.  We write this as $P(x\rightarrow x')$.  The uniformity of the lattice reduces us to studying $P(x\rightarrow 0)$.  As in \cite[Chapter 2]{OV}, we have for sufficiently small $\sigma^2$ that
\begin{equation}\label{first_pep}
P_e\leq c\sum_{0\neq x \in \mathcal{C}} P(x\rightarrow 0) \leq d\sum_{0\neq x\in \mathcal{C}}\prod_{i = 1}^n\frac{1}{|x_i|^2} = d\sum_{\substack{0\neq x\in\Lambda \\ ||x||\leq R}}\prod_{i = 1}^n\frac{1}{|x_i|^2}
\end{equation}
where $c$ and $d$ depend on the noise variance $\sigma^2$ and the dimension $n$, but not $\Lambda$.  Here we have implicitly assumed that $x_i\neq 0$ for all $x\neq 0$ and all $i$, which is ultimately true of the ideal lattices we consider.  Thus inverse norm sums show up in the context of the PEP.

\subsection{The wiretap channel and the probability of Eve's correct decision}
In a wiretap channel, Alice is transmitting confidential data to the intended receiver Bob over a Rayleigh fading channel, while an eavesdropper Eve tries to intercept the data received over another Rayleigh fading channel. The security is based on the assumption that Bob's SNR is sufficiently large compared to Eve's SNR. In addition, a coset coding strategy \cite{Wyner} is employed to confuse Eve.  We assume both Bob and Eve have perfect channel state information, while Alice has none. The details of the channel model and related probability expressions can be found in \cite{BO_wiretap}.

In coset coding, random bits are transmitted in addition to the data bits.  Let us denote the lattice intended for Bob by $\Lambda_b$, and by $\Lambda_e\subset\Lambda_b$ the sublattice encoding the random bits intended for Eve's confusion. Now the transmitted codeword $x$ is picked from a coset $\Lambda_e+c$ belonging to the disjoint union 
\begin{equation}
\Lambda_b=\cup_{j=1}^{2^k} \Lambda_e+c_j
\end{equation}
encoding $k$ bits:
\begin{equation}
x=r+c\in \Lambda_e+c,
\end{equation}
where $\textbf{r}$ encodes the random bits, and $\textbf{c}$ contains the data bits.

Next, let us recall the expression $P_{c,e}$ of the probability of a correct decision for Eve, when observing a lattice $\Lambda_e$ and having large enough SNR for decoding $\Lambda_e$.  For the fast fading case \cite[Sec. III-A]{BO_wiretap},
\begin{equation}\label{fast-prob}
P_{c,e}\approx \left(\frac 1{4\gamma_e^2}\right)^{n/2}\textrm{Vol}(\Lambda_b) \sum_{\substack{0\neq x \in\Lambda_e \\ ||x||\leq R}}\prod_{i=1}^n\frac{1}{|x_i|^3},
\end{equation}
where $\gamma_e$ is the average SNR for Eve assumed sufficiently large so that Eve can perfectly decode $\Lambda_e$.  It can be concluded that the smaller the sum is in (\ref{fast-prob}) the more confusion Eve is experiencing.  Here we have implicitly assumed that $x_i\neq0$ for all $x$, which will ultimately be true of the full-diversity ideal lattices we use.

\subsection{Inverse Norm Sums of Ideal Lattices}

We now restrict our number field $K$ to be either totally real of degree $n$ over $\Q$, with distinct embeddings $\sigma_1,\ldots,\sigma_n$ into $\R$.  The restriction to totally real number guarantees full diversity and also conveniently forces a relation between the product distance and the algebraic norm.   We also restrict from now on to $||\cdot|| = ||\cdot||_{\infty}$, so that $||x||_{\infty} = \max_i |x_i|$, and our constellations $\Lambda\cap \{x\in \R^n\ |\ ||x||_{\infty}\leq R\}$ are the points in $\Lambda$ inside a box of side length $2R$ centered at the origin.  This restriction is mostly for convenience as it makes proving our Theorem \ref{bkR_estimate} easier.  However, any norm of the form $||\cdot ||_p$ can be used, so that our results also apply to, for example, spherically shaped constellations.

The authors of \cite{BO_wiretap} propose using an ideal lattice from a totally real number field $K$ as Eve's lattice.   The resulting sums from the previous section can then be analyzed using number theoretic methods.  Additionally, carefully chosen ideal lattices are known to give Bob good performance.    Suppose now that Alice and Bob employ coset coding to confuse Eve with $\Lambda_e = \Lambda = (\frak{a},q_\alpha)$ an ideal lattice, scaled by a constant $\kappa$ so that $\vol(\Lambda) = 1$.  The corresponding probability of Eve's correct decision \eqref{fast-prob} yields the following inverse norm sum (cf. \cite[Sec. III-B]{BO_wiretap} for the original form of this sum):
\begin{equation}\label{first_ins}
\boxed{S(\Lambda,s,R) = \sum_{\substack{0\neq x\in \frak{a} \\ ||\kappa \psi_\alpha(x)||_{\infty}\leq R}} \prod_{i = 1}^n\frac{1}{|\kappa\psi_\alpha(x)_i|^s} 
= \frac{1}{\kappa^{ns}|N(\alpha)|^{s/2}}\sum_{\substack{0\neq x\in \frak{a} \\ ||\psi_\alpha(x)||_{\infty}\leq R/\kappa}} \frac{1}{|N(x)|^s} }
\end{equation}
which is our main object of study.  The use of the variable $s$ in (\ref{first_ins}) allows us to simultaneously analyze the cases of $s = 2$ (the pairwise error probability for the Rayleigh fading channel) and $s = 3$ (Eve's probability of correct decision).  Without a bound  on $||\cdot||_{\infty}$, the sum (\ref{first_ins}) is infinite except in the special case of $K = \Q$ or $K$ an imaginary quadratic field,  which are of limited interest to applications. 

\section{First observations and bounds}
\label{bounds1}

To establish some simple bounds for inverse norm sums, let us first consider an ideal $\frak{a}\subseteq\mathcal{O}_K$ in a totally real number field $K$ of degree $n$ over $\Q$.  We consider its (untwisted) canonical embedding $\psi:\frak{a}\rightarrow \R^n$ and the corresponding lattice $\Lambda^0 = \psi(\frak{a})$.  The inverse norm sum we are interested in for this section is
\begin{equation}
S(\Lambda^0,s,R) = \sum_{\substack{0\neq x\in \frak{a} \\ ||\psi(x)||_{\infty}\leq R}}\frac{1}{|N(x)|^s} = \sum_{k = 1}^{R^n}\frac{b_{k,R}^{\frak{a}}}{k^s}
\end{equation}
where
\begin{equation}
b_{k,R}^{\frak{a}} = \#\{x\in \frak{a}\ |\ |N(x)| = k\ \text{and}\ ||\psi(x)||_{\infty}\leq R\}
\end{equation}
and we note that clearly $b_{k,R}^{\frak{a}} = 0$ for $k> R^n$.  Albeit straightforward, the following result gives us a nontrivial lower and upper bound for the sum $S(\Lambda^0,s,R)$. Notice that below we have not normalized the lattice to have unit volume.

\begin{prop}
Let $\Lambda^0 = (\frak{a},q_1)$ be an (untwisted, unnormalized) ideal lattice, let $m$ be the order of $[\frak{a}]$ in the class group $C_K$ of $K$, let $N = N(\frak{a})$, and let $M_R = \max_k\{b_{k,R}\ |\ k\leq R^n\}$.  Then for sufficiently large $R$ we have
\begin{equation}
\boxed{
\frac{b_{N^m,R}^{\frak{a}}}{N^{ms}}\leq S(\Lambda^0,s,R)\leq M_R\zeta(s)
}
\end{equation}
where $\zeta(s) = \sum_{k\geq 1}1/k^s$ is the familiar Riemann zeta function.
\end{prop}
\begin{IEEEproof}
Let us start with the lower bound.  Since $m$ is the order of $\frak{a}$ in the ideal class group, we must have that $\frak{a}^m = (\alpha)$ for some $\alpha\in\mathcal{O}_K$.  Then $|N(\alpha)| = N^m$ by multiplicativity of the norm.  Choose $R$ sufficiently large so that
\begin{equation}
\{x\in (\alpha)\ |\ x \text{ generates } (\alpha) \text{ and }||\psi(x)||_{\infty}\leq R\}\neq \emptyset
\end{equation}
so that $b_{N^m,R}^{\frak{a}}\neq 0$.  The lower bound follows easily.  For the upper bound, a simple computation gives us
\begin{equation}
S(\Lambda^0,s,R)=\sum_{k = 1}^{R^n}\frac{b_{k,R}^{\frak{a}}}{k^{s}}
\leq M_R\sum_{k = 1}^{R^n}\frac{1}{k^s}
\leq M_R\zeta(s).
\end{equation}
which completes the proof.
\end{IEEEproof}

When $\frak{a} = \mathcal{O}_K$ then of course $m = 1$ and it suffices to take $R\geq 1$.  The lower bound then reduces to the number of units in the bounding box.  These first simple bounds are not very tight. Our goal in the next section is to derive more precise estimates of $b_{k,R}^{\frak{a}}$ arising from geometric analysis.  These estimates will ultimately be combined to estimate the full inverse norm sum, for twisted, normalized lattices.

\section{Estimating the quantity $b_{k,R}^{\frak{a}}$}
\label{geometric}

In this section we fix $K$ be a totally real number field of degree $n$ over $\Q$, an ideal $\frak{a}\subseteq\mathcal{O}_K$, and its canonical embedding $\psi:\frak{a}\rightarrow \R^n$, without any twisting element.  The main result in this section is Theorem \ref{bkR_estimate} which provides an estimate to
\begin{equation}\label{bkra_defn}
b_{k,R}^{\frak{a}} = \#\{x\in \frak{a}\ |\ |N(x)| = k \text{ and } H(x)\leq R\}.
\end{equation}
Before estimating the quantity $b_{k,R}^{\frak{a}}$ we first prove the following lemma, which allows us to count principal ideals of a given norm contained in a given ideal.  For any ideal $\frak{a}\subseteq\mathcal{O}_K$ and any ideal class $[\frak{b}]$, we define
\begin{align}
a_{k}^{[1],\frak{a}} &= \#\{(\alpha)\subseteq\frak{a}\ |\ |N(\alpha)| = k \} \\
a_k^{[\frak{b}]} &= \#\{\frak{c}\subseteq\mathcal{O}_K\ |\ N(\frak{c}) = k \text{ and } [\frak{c}] = [\frak{b}]\}
\end{align}
for $k>0$.  The following lemma relates these two quantities, and actually does not depend on $K$ being totally real.

\begin{lemma}\label{prinideal_coeffs}
Let $K$ be a number field, let $\frak{a}\subseteq\mathcal{O}_K$ be an ideal with norm $N = N(\frak{a})$, and let $[\frak{a}]^{-1} = [\frak{a}]$ be the inverse of the class of $\frak{a}$ in the ideal class group $C_K$ of $K$.  Then
\begin{equation}
\boxed{a^{[1],\frak{a}}_{kN} = a_k^{[\frak{a}]^{-1}}}
\end{equation}
for any $k>0$.
\end{lemma}
\begin{IEEEproof}
Let $A$ be the set of all ideals of $\mathcal{O}_K$, and let $A^{\frak{a}}$ be the set of all ideals which are contained in $\frak{a}$.  Then we claim that the map
\begin{equation}
\phi_{\frak{a}}: A\rightarrow A^\frak{a},\quad \phi(\frak{c}) = \frak{a}\frak{c}
\end{equation}
is a bijection.  Indeed, we can define an inverse $\psi_{\frak{a}}:A^\frak{a}\rightarrow A$ in the following way.  If $\frak{c}'\subseteq\frak{a}$ then by basic properties of Dedekind domains there must exist an ideal $\frak{c}$ so that $\frak{c}' = \frak{a}\frak{c}$.  The ideal $\frak{c}$ is unique by, for example, prime factorization.  Now define $\psi_\frak{a}(\frak{c}') = \frak{c}$, and it is easy to check that $\phi_\frak{a}\circ \psi_\frak{a}$ and $\psi_{\frak{a}} \circ \phi_{\frak{a}}$ are both the identity map.

We see that $\phi_{\frak{a}}$ multiplies norms of ideals by $N$ in the following sense: 
\begin{equation}
N(\phi_\frak{a}(\frak{c})) = N(\frak{a})N(\frak{c}) = NN(\frak{c})
\end{equation}
and hence induces bijection between ideals of norm $k$ and ideals of norm $kN$ which are contained in $\frak{a}$.  Now for fixed $k_1,k_2>0$ and some ideal classes $[\frak{c}]$ and $[\frak{d}]$, and define
\begin{equation}
A_{k_1}^{[\frak{c}]} := \{\frak{c}' \subseteq\mathcal{O}_K\ |\ N(\frak{c}') = k_1 \text{ and }[\frak{c}'] = [\frak{c}]\}
\quad
\text{and}
\quad
A_{k_2}^{[\frak{d}],\frak{a}} := \{\frak{d}' \subseteq\frak{a}\ |\ N(\frak{d}') = k_2 \text{ and }[\frak{d}'] = [\frak{d}]\}.
\end{equation}
Then it is clear that for any ideal class $[\frak{c}]$ the function $\phi_\frak{a}$ induces a bijection
\begin{equation}
\phi_\frak{a}:A_k^{[\frak{c}]}\rightarrow A_{kN}^{[\frak{ac}],\frak{a}}
\end{equation}
Setting $[\frak{c}] = [\frak{a}]^{-1}$ to be the inverse of $[\frak{a}]$ in the ideal class group completes the proof, since $a_k^{[\frak{a}]^{-1}} = \# A_k^{[\frak{a}]^{-1}}$ and $a_{kN}^{[1],\frak{a}} = \#A_{kN}^{[1],\frak{a}}$.
\end{IEEEproof}

We remark that if $(\alpha)\subseteq\frak{a}$ then by basic properties of Dedekind domains, we have $\frak{a}|(\alpha)$.  Taking norms gives us that $N(\frak{a})|N(\alpha)$ as integers.  Hence the norm of any principal ideal contained in $\frak{a}$ must be a multiple of $N(\frak{a})$, and so the above lemma does indeed count all possible principal ideals contained in $\frak{a}$.

Since $K$ is totally real we of course have $w_K = 2$.  However, to suggestively hint at a possible connection with the Class Number Formula and generalizations to $K$ which are not totally real, we write $w_K$ in the following theorem.  One could use the above lemma to rewrite the following theorem in terms of the Dirichlet coefficients $a_{kN}^{[\frak{a}]^{-1}}$, but the given incarnation appears more streamlined.

\begin{theorem}\label{bkR_estimate}
Let $K$ be a totally real number field of degree $n$ over $\Q$, and consider the canonical embedding (cf.\ (\ref{can-emb})) $\psi:\frak{a}\rightarrow \R^n$ of an ideal $\frak{a}\subseteq\mathcal{O}_K$.   Let $b_{k,R}^\frak{a}$ be defined as in (\ref{bkra_defn}).  Then
\begin{equation}\label{term_approx}
\boxed{b_{k,R}^{\frak{a}} = \frac{w_K}{R_K(n-1)!} a_{k}^{[1],\frak{a}}\log(R^n/k)^{n-1} + O(\log(R^n/k)^{n-2})}
\end{equation}
as $R\rightarrow \infty$, that is, as the size of the constellation increases.  \end{theorem}
\begin{IEEEproof}

Let us define the set
\[
Z_k := \left\{(x_1,\ldots,x_n)\ |\ \prod_{i=1}^n |x_i| = k\right\} \subset \R^n
\]
so that the canonical embedding induces a bijection
\begin{equation}\label{bij}
\psi: \{x\in \frak{a}\ |\ |N(x)| = k\} \rightarrow \psi(\frak{a})\cap Z_k
\end{equation}
To count the elements of height bounded by $R$ on the left-hand side of (\ref{bij}) we will work instead with the more ``geometric'' right-hand side.  Let us define the logarithm map $\log:\R^n\rightarrow \R^n$ by
\[
\log(x_1,\ldots,x_n) = (X_1,\ldots,X_n),\quad X_i = \log|x_i|
\]
The logarithm map linearizes the sets $Z_k$ by taking them to hyperplanes:
\[
\log(Z_k) = \Hc_k := \{(X_1,\ldots,X_n)\ |\ X_1 + \cdots + X_n = \log(k)\}
\]
Furthermore, we have $\log(\psi(x)) = \log(\psi(y))$ for $x,y\in\frak{a}$ if and only if there exists a root of unity $\zeta\in\mathcal{O}_K^\times$ such that $x = \zeta y$.  Therefore when restricted to $\psi(\frak{a})\cap Z_k$, the logarithm is $w_K$-to-$1$, where we recall that $w_K$ is the number of roots of unity in $K$.

To see what happens to vectors of bounded height under the logarithm map, we note that the bounding box $\mathcal{B}_R$ is transformed into the semi-infinite rectangular region
\begin{equation}
\log(\mathcal{B}_R) = (-\infty,\log(R)]^n
\end{equation}
which has a single vertex at $(\log(R),\ldots, \log(R))$. Denote the intersection of the hyperplane $\mathcal{H}_k$ with $\log(\mathcal{B}_R)$ by
\begin{equation}
\mathcal{S}_k :=\log(\mathcal{B}_R) \cap \Hc_k.
\end{equation}
Note that this is nonempty exactly when $1\leq k \leq R^n$.  Taking the logarithm map has essentially reduced our problem to counting the number of lattice points which are in $\mathcal{S}_k$ after the logarithm map.  This requires knowing the volume of $\mathcal{S}_k$, which we can compute as follows.  Observe that $\mathcal{S}_k$ is the basis of a hyper-pyramid $V_k$ with a vertex at $(\log(R),\ldots, \log(R))$, whose volume is equal to the volume of a simplex with $n$ orthogonal vectors of length 
$n \log(R)- \log(k)$, i.e.,
\begin{equation}
\vol(V_k) = \frac{(n \log(R)- \log(k))^n}{n!} = \frac{\log(R^n/k)^n}{n!}.
\end{equation}
The height of $V_k$ is given by $\text{ht}(V_k) = (n \log(R)- \log(k))/\sqrt{n} = \log(R^n/k)/\sqrt{n}$, hence
\begin{equation}
\vol(\mathcal{S}_k) = n\frac{\vol(V_k)}{\text{ht}(V_k)}  = \frac{\sqrt{n}}{(n-1)!}\log(R^n/k)^{n-1}.
\end{equation}


Let us, for starters, suppose that $\frak{a} = \mathcal{O}_K$ and that $k = 1$, which reduces us to counting the number of units in $\mathcal{B}_R$.  By the Dirichlet Unit Theorem, the units form a lattice under the logarithm map:
\begin{equation}
\Lambda_{\log}:=\log(\psi(\mathcal{O}_K^\times)) \subset \mathcal{H}_1,\quad \vol(\Lambda_{\log}) = R_K\sqrt{n}
\end{equation}
where we recall that $R_K$ is the regulator of $K$.  Since the logarithm map is $w_K$-to-$1$, we can estimate the number of units in $\mathcal{B}_R$ by dividing the volume of $\mathcal{S}_k$ by the volume of $\Lambda_{\log}$, as in \cite[Chapter VI \S2, Theorem 2]{lang}:
\begin{equation}
b_{1,R} = w_K\frac{\vol(\mathcal{S}_1)}{ \vol(\Lambda_{\log})} + O(\log(R^n)^{n-2}) = \frac{w_K}{R_K(n-1)!} \log(R^n)^{n-1}+ O(\log(R^n)^{n-2})
\end{equation}
This proves the theorem for units, i.e.\ when $\frak{a} = \mathcal{O}_K$ and $k = 1$.

For non-units ($k>1$) and proper ideals $\frak{a}\subsetneq \mathcal{O}_K$ the problem is more complicated.  Since $|N(\alpha u)| = |N(\alpha)|$ for all units $u$  and the norm of a principal ideal is equal to the absolute norm of any generator we can conclude that for $k>1$, $\log(\psi(\frak{a})\cap Z_k)$ is a union of exactly $a_k^{[1],\frak{a}}$ translates of $\Lambda_{\log}$. Then we can estimate $b_{k,R}^{\frak{a}}$ by 
\begin{align}\label{err_fn}
b_{k,R}^{\frak{a}} &= w_K a_k^{[1],\frak{a}}\frac{ \vol(\mathcal{S}_k)}{ \vol(\Lambda_{\log})} +O(\log(R^n/k)^{n-2}) \\
&= \frac{w_K}{R_K(n-1)!} a_k^{[1],\frak{a}}\log(R^n/k)^{n-1} + O(\log(R^n/k)^{n-2})
\end{align}
as desired.
\end{IEEEproof}

\begin{figure}[h!]
\begin{center}
\includegraphics[width=.4\textwidth]{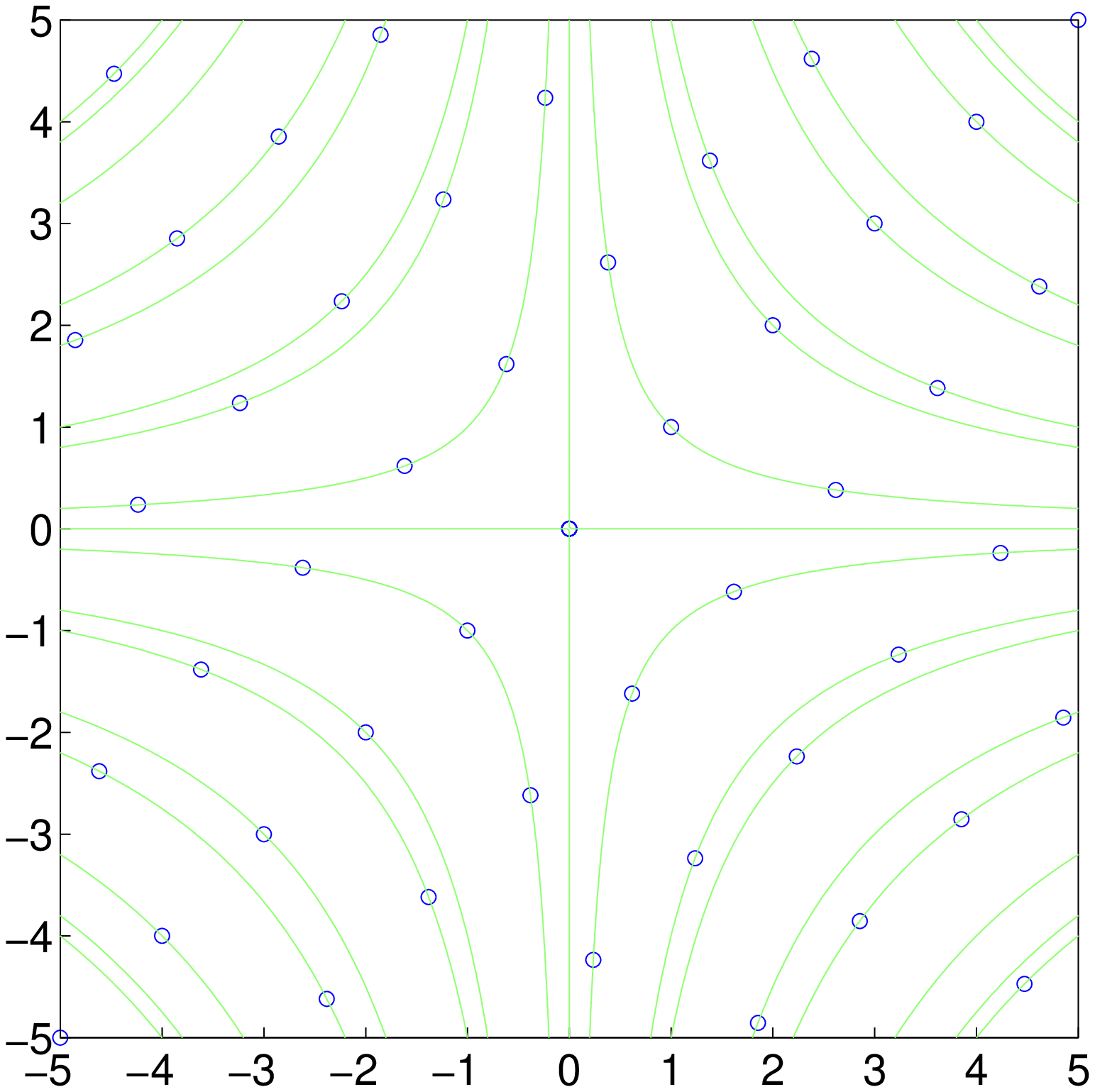}
\hspace{1cm}
\includegraphics[width=.385\textwidth]{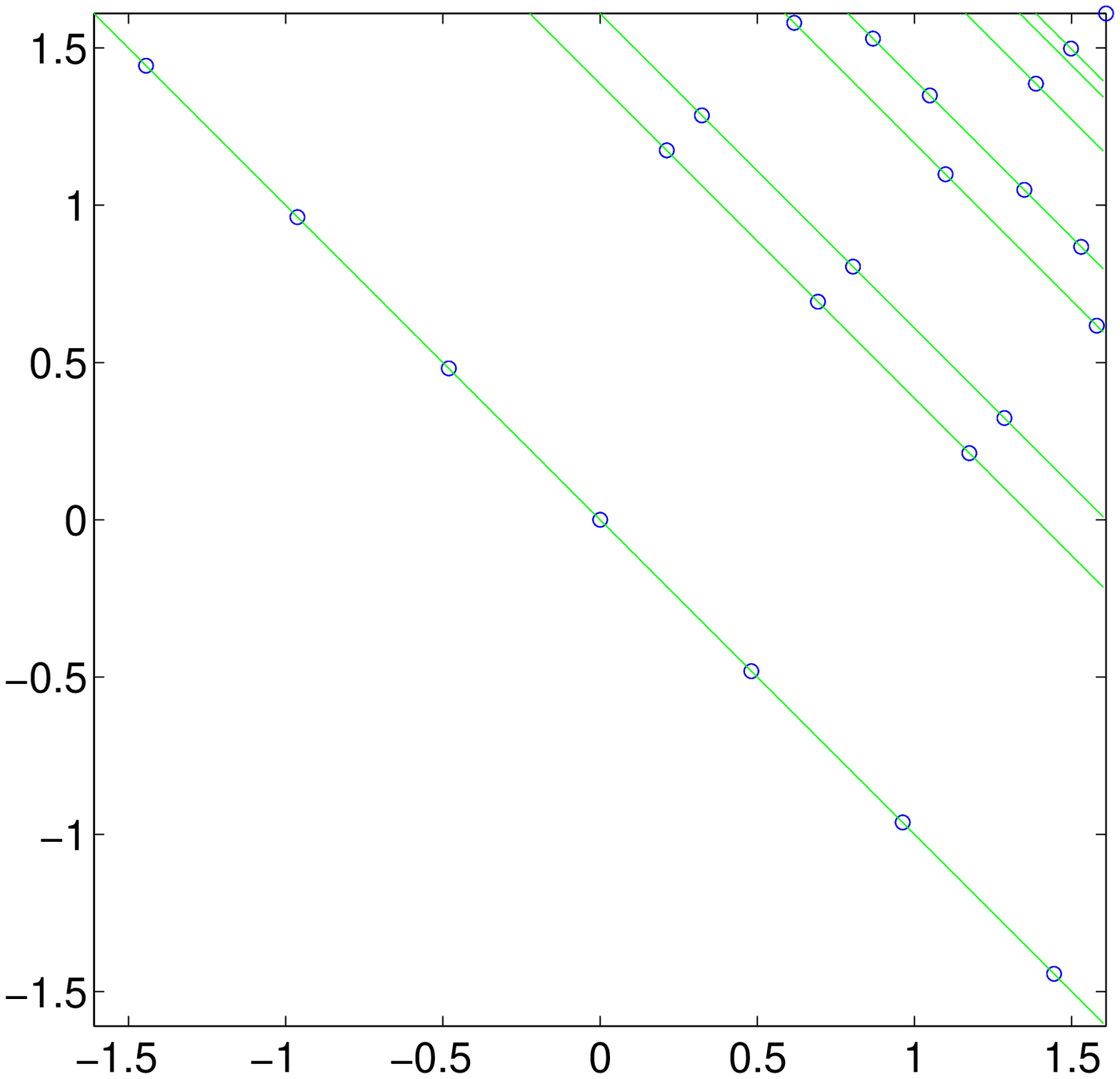}
\end{center}
\caption{On the left, the canonical embedding of the ideal $\frak{a}=\mathcal{O}_K$ of $K = \Q(\sqrt{5})$ with $R = 5$.  On the right, its image under the logarithm map.  The green hyperbolas in the left figure, i.e.\ the $Z_k$, have been taken to the green hyperplanes $\mathcal{H}_k$ in the right figure.}
\label{Fig2D}
\end{figure}

To illustrate the accuracy of our estimation, let us consider some example cases in more detail.  In the following two examples, the fields we consider satisfy $h_K = 1$ and we consider the lattice defined by $\frak{a} = \mathcal{O}_K$.  Hence out of convenience we drop the superscripts on the Dirichlet coefficients, and define the following:
\begin{equation}
n_{k,R} = \frac{w_K}{R_K(n-1)!} a_{k}\log(R^n/k)^{n-1},\quad f_{k,R} = \left\lfloor |n_{k,R} - b_{k,R}|\right\rfloor
\end{equation}
so that $f_{k,R}$ measures the accuracy of our approximation.  The error function $f_{k,R}$ grows quite large when the dimension of the lattice grows. We will illustrate the size of the error function in the following example.

\begin{example}
We start with the field $K=\mathbf{Q}(\sqrt{5})$, see Fig. \ref{Fig2D} for the illustration of the lattice and the logarithmic lattice.   Let us first set $R=10$, i.e., $1\leq k\leq 100$.  The values of $n_{k,R}$, $b_{k,R}$, and $f_{k,R}$ (the length of the segment connecting the previous two) are collected in Fig.\ \ref{actual_and_est_and_error}. We can see that the error satisfies $f_{k,R} \leq 2$ for all $k$. The values are only given for those $k$ for which $a_k\neq 0$, that is, there exists a principal ideal of norm $k$. For all other $k$ we have $b_{k,R}=f_{k,R}=0$.  When we increase the size of the constellation by considering norms up to $k=2000$, i.e., $R=\sqrt{2000}$, we still have $f_{k,R}\leq 3$ for all $k$, see Fig.\ \ref{actual_and_est_and_error}.

In Fig.\ \ref{actual_and_est} we separately plot the actual values of $b_{k,R}$ and the estimates $n_{k,R}$, to emphasize that the error in such an approximation is unavoidable.  Essentially, we are approximating a staircase function with a smooth function.

\begin{figure}[h!]
\begin{center}
\includegraphics[width=.48\textwidth]{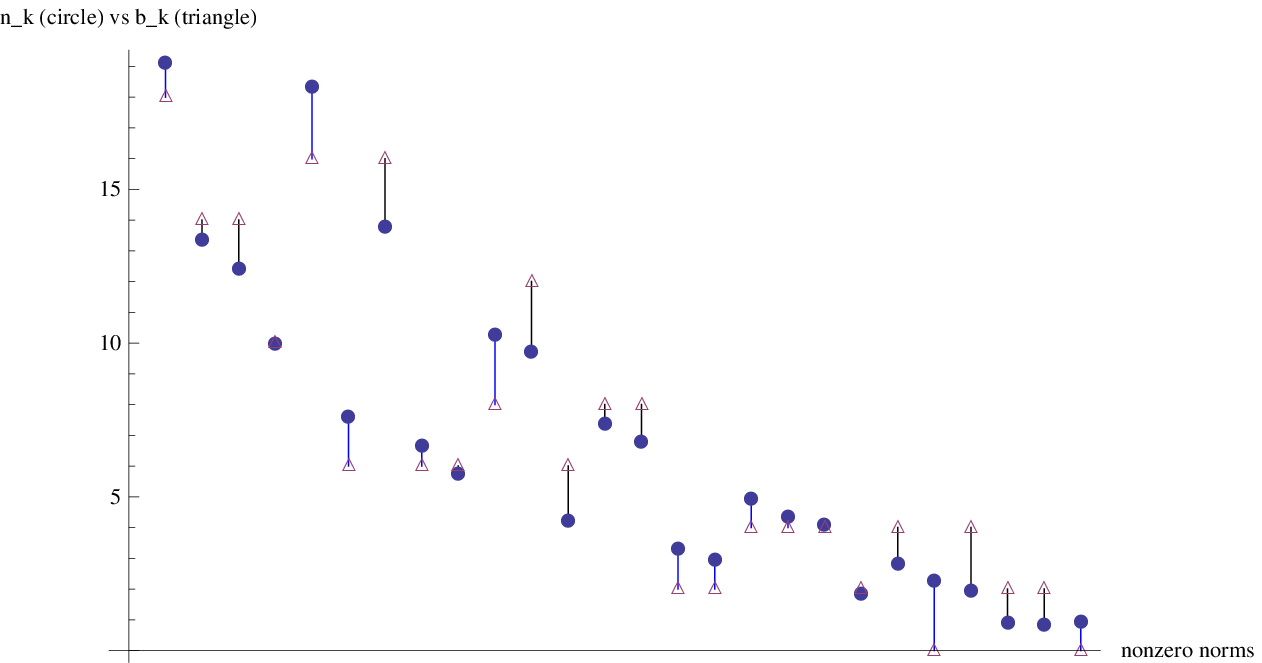}
\includegraphics[width=.48\textwidth]{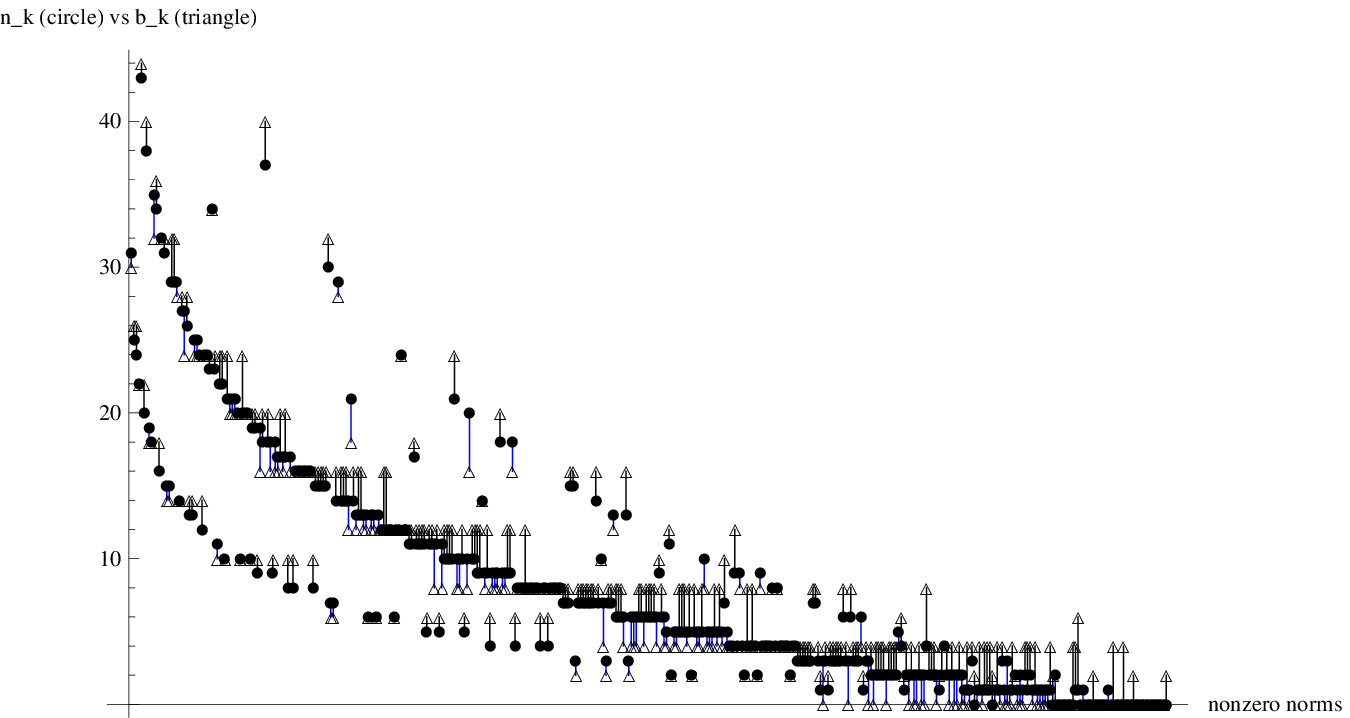}
\end{center}
\caption{The estimates $n_{k,R}$ (circles) and the exact values $b_{k,R}$ (triangles) for the ring of integers of $K = \Q(\sqrt{5})$.  On the left we have $1\leq k\leq R^2 = 100$, and on the right we have extended to $1\leq k\leq R^2 = 2000$. }
\label{actual_and_est_and_error}
\end{figure}

\begin{figure}[h!]
\begin{center}
\includegraphics[width=.45\textwidth]{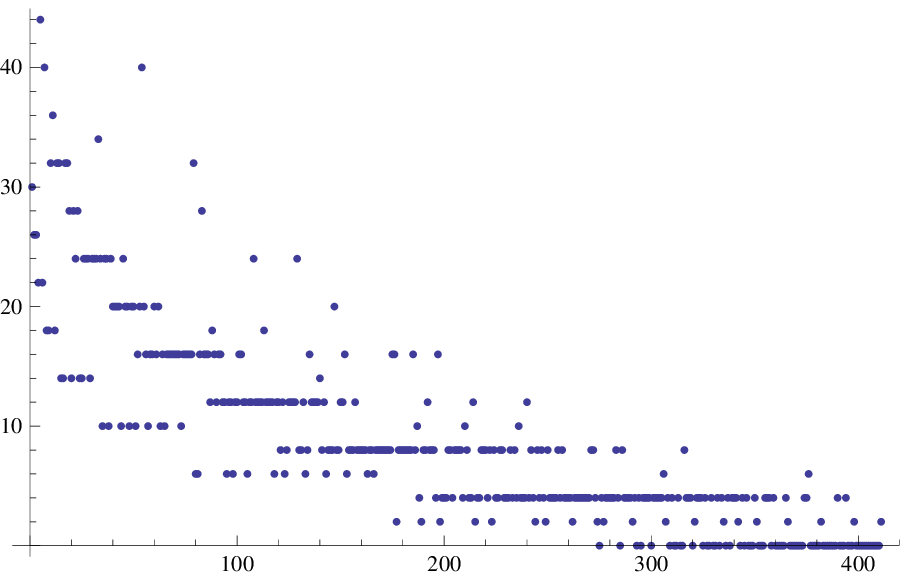}
\includegraphics[width=.45\textwidth]{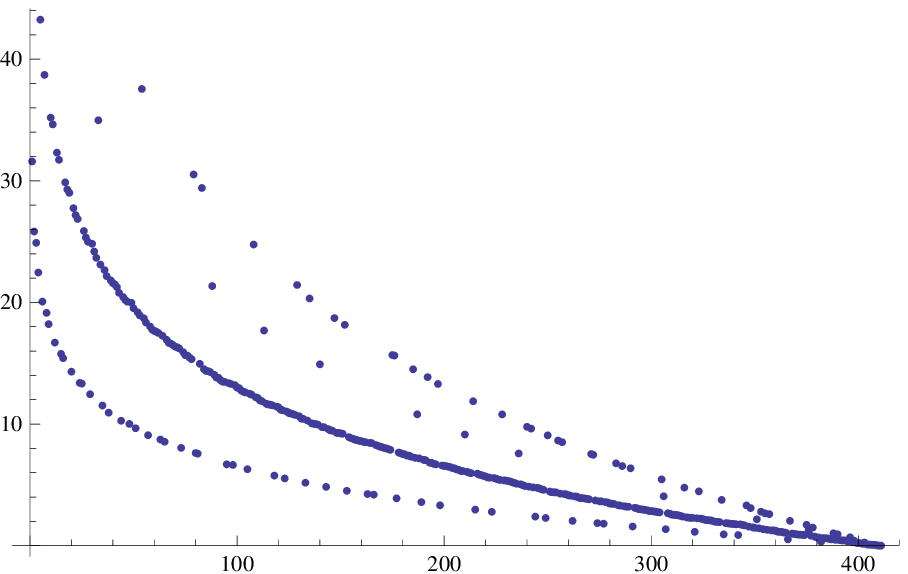}
\end{center}
\caption{The exact values $b_{k,R}$ on the left, and the estimates $n_{k,R}$ on the right, for the canonical embedding of the ring of integers of $K = \Q(\sqrt{5})$.  The different ``curves'' swept out on the right correspond to the different values of $a_k$, and the apparent continuity comes from the term $\log(R^n/k)^{n-1}$.  }
\label{actual_and_est}
\end{figure}

\end{example}

\begin{figure}[h!]
\begin{center}
\includegraphics[width=.45\textwidth]{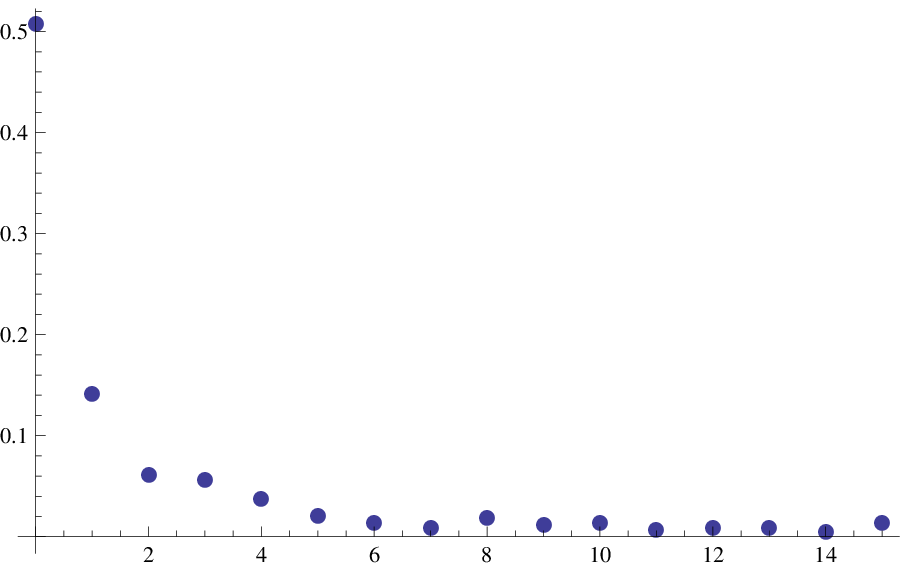} 
\includegraphics[width=.45\textwidth]{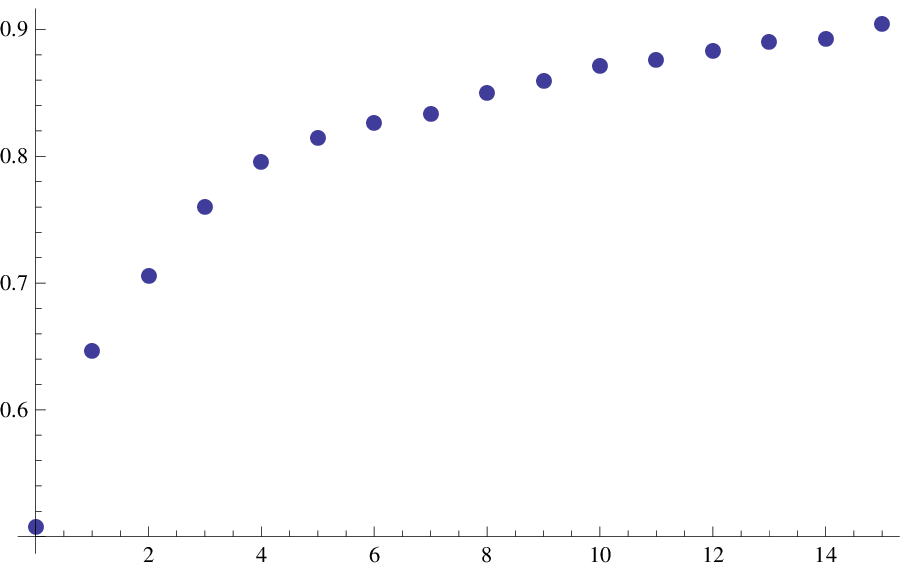}
\end{center}
\caption{The frequency (left) and cumulative frequency (right) of estimation errors as a function of $k$, $1\leq k \leq 65536$ for the field $K = \Q(\zeta_{32} + \zeta_{32}^{-1})$ with $n = 8$. The edge length of the bounding hypercube is $2R=10$.}
\label{n8error}
\end{figure}

\begin{example}
In order to see what happens to the size of error $f_{k,R}$ when the dimension grows, let us consider a case with $n=8$. This is already quite a high delay in practice, as we require encoding over eight time instances.   The field $K$ is the maximal totally real subfield of the $32^{nd}$ cyclotomic field, $K=\Q(\zeta_{32}+\zeta_{32}^{-1})$.

While the absolute error increases with the dimension, it is still negligible considering that out of all $k$ considered more than half satisfy $n_{k,R}=b_{k,R}$, meaning no error.   For the rest of the cases (meaning an error occurs) either the error is very small, or (a bigger error) occurs very rarely.  In Fig.\ \ref{n8error} we have depicted the frequency and cumulative frequency of errors, respectively, as a function of $k$. One can see that  cumulative frequency as high as 90\% is achieved already by errors of size  $\leq 15$.
\end{example}

\section{Approximating the Inverse Norm Sum}
\label{main_section}

The goal of this section is to use the above estimate of $b_{k,R}^{\frak{a}}$ to estimate $S(\Lambda,s,R)$ and prove Theorem \ref{main_theorem}.  Understanding the error term in such an approximation will ultimately depend on bounding the tail of the derivatives of the zeta functions in question, which we do in the following lemma.  Let us write the $m^{th}$ derivative of an ideal class Dedekind zeta function of our number field $K$ as
\begin{align}
\zeta_K^{(m),{[\frak{a}]}}(s) &= \sum_{ k = 1}^{\infty} (-1)^m\frac{a^{[\frak{a}]}_k\log(k)^m}{k^s} \\
&= \sum_{ k = 1}^{R^n} (-1)^m\frac{a^{[\frak{a}]}_k\log(k)^m}{k^s} + \sum_{ k = R^n + 1}^{\infty} (-1)^m\frac{a^{[\frak{a}]}_k\log(k)^m}{k^s}
\end{align}
The proof of our main theorem will require us to bound the absolute value of the tail of the ideal class zeta function, which our next lemma accomplishes.
\begin{lemma}\label{ekaarvio}Suppose that $R \geq 3$, let $[\frak{a}]$ be an ideal class in $K$, and let $N$ be a constant.  We have
\begin{equation}
\sum_{k= (R^n +1)/N}^{\infty}\frac{a^{[\frak{a}]}_k(\log (kN))^m}{k^s}\leq \begin{cases}c R^{-n}(\log (R^n))^m,\quad&\textrm{when $s=2$}\\ cR^{-2n}(\log (R^n))^m,\quad & \textrm{when $s=3$}\end{cases}
\end{equation}
where $c$ is a constant depending on the field $K$ and the ideal $\mathfrak{a}$, but not on $R$.

\end{lemma}
\begin{IEEEproof}
We relegate the proof to the Appendix.
\end{IEEEproof}
This lemma is useful in that compared to the approximate size of the inverse norm sum, the tails of the ideal class Dedekind zeta functions are quite small.  Thus the error introduced by including or excluding the tails of the zeta functions does not affect the growth of the inverse norm sum.

We are ready to state and prove the main theorem of the paper.  Let $K/\Q$ be a totally real number field of degree $n$, and let $\Lambda = (\frak{a},q_\alpha)$ be an ideal lattice with twisted canonical embedding $\psi_\alpha:\frak{a}\rightarrow\R^n$, scaled by a constant $\kappa$ so that $\vol(\Lambda) = 1$.  We consider a finite constellation
\begin{equation}
\Lambda \cap \mathcal{B}_R,\quad \text{where}\quad \mathcal{B}_R := \{ x\in \R^n\ |\ ||x||_{\infty}\leq R\}
\end{equation} 
so that the bounding region is a hypercube of side length $2R$ centered at the origin.   Recall the corresponding inverse norm sum
\begin{equation}
S(\Lambda,s,R) = \sum_{\substack{0\neq x\in\frak{a} \\ ||\kappa\psi_\alpha(x)||_{\infty}\leq R}} \prod_{i = 1}^n\frac{1}{|\kappa\psi_\alpha(x)_i|^s} = \frac{1}{\kappa^{ns}|N(\alpha)|^{s/2}}\sum_{\substack{0\neq x\in\frak{a} \\ ||\psi_\alpha(x)||_{\infty}\leq R/\kappa}} \frac{1}{|N(x)|^s} 
\end{equation}
which was defined in (\ref{first_ins}).  Theorem \ref{main_theorem} describes this inverse norm sum as a function of the bound $R$.

\begin{theorem}\label{main_theorem}
Let $K/\Q$ be a totally real number field of degree $n$, let $\Lambda = (\frak{a},q_\alpha)$ be an ideal lattice with twisted canonical embedding $\psi_\alpha$, scaled by $\kappa$ so that $\vol(\Lambda) = 1$.  Let $[\frak{a}]^{-1}$ be the inverse of the class of $[\frak{a}]$ in the ideal class group.  Then the inverse norm sum $S(\Lambda,s,R)$ satisfies
\begin{equation}
\boxed{S(\Lambda,s,R) = \frac{w_K|D_K|^{s/2}}{R_K}\zeta_K^{[\frak{a}]^{-1}}(s)c_n\log(R)^{n-1} + O(\log(R)^{n-2})}
\end{equation}
where $c_n = n^{n-1}/(n-1)!$ depends only on $n$.
\end{theorem}

\begin{IEEEproof}
To use the estimate of $b_{k,R}^{\frak{a}}$ in Theorem \ref{bkR_estimate} we need to consider the unscaled, untwisted canonical embedding of $\frak{a}$, which we can reduce to as follows.  The inverse norm sum $S(\Lambda,s,R)$ appears to depend on the twisting element $\alpha$ and the constant $\kappa$, but we can essentially remove this dependence.  Define the constants $m^0_\alpha = \min_i |\sqrt{\sigma_i(\alpha)}|$ and $m^1_\alpha = \max_i |\sqrt{\sigma_i(\alpha)}|$, and let $\psi:\frak{a}\rightarrow \R^n$ denote the canonical embedding (with twisting element $\alpha = 1$ and no scaling).  It is then straightforward to show that 
\begin{equation}\label{three_sums}
\sum_{\substack{0\neq x\in \frak{a}\\ ||\psi(x)||_{\infty}\leq R/(\kappa m^0_\alpha)}}\frac{1}{|N(x)|^s}
\leq 
\sum_{\substack{0\neq x\in \frak{a}\\ ||\psi_\alpha(x)||_{\infty}\leq R/\kappa}}\frac{1}{|N(x)|^s}
\leq
\sum_{\substack{0\neq x\in \frak{a}\\ ||\psi(x)||_{\infty}\leq R/(\kappa m^1_\alpha)}}\frac{1}{|N(x)|^s}
\end{equation}
If $c>0$ is any constant, we can use simple binomial expansion to show that
\begin{align}
\log(R/c)^{n-1} &= (\log(R) - \log(c))^{n-1} \\
& = \sum_{m = 0}\binom{n-1}{m}\log(R)^{n-1-m}\log(c)^m \\
& = \log(R)^{n-1} + O(\log(R)^{n-2}
\end{align}
Let $\Lambda^0 = (\frak{a},q_1)$ denote the unscaled lattice corresponding to the untwisted canonical embedding $\psi$.  Up to the multiplicative constant $\kappa^{ns}|N(\alpha)|^{s/2}$ and an additive error term which is of the order $O(\log(R)^{n-2})$, all three of the sums in (\ref{three_sums}) will have the same behavior as
\begin{equation}
S(\Lambda^0,s,R)=\sum_{\substack{0\neq x\in\frak{a} \\ ||\psi(x)||_{\infty}\leq R}}\frac{1}{|N(x)|^s}=\sum_{k=1}^{R^n}\frac {b_{k,R}^{\frak{a}}}{k^s},
\end{equation}
for sufficiently large $R$, where we note that $b_{k,R}^{\frak{a}} = 0$ if $k> R^n$.  

If $\Lambda'$ denotes the \emph{unscaled} ideal lattice defined by $(\frak{a},q_\alpha)$, then $\kappa\Lambda' = \Lambda$ and it follows that $1 = \vol(\kappa\Lambda') = \kappa^n\vol(\Lambda')$ and hence $\kappa = \vol(\Lambda')^{-1/n}$.  Since $\vol(\Lambda')^2 = |N(\alpha)|N(\frak{a})^2|D_K|$ (see \cite[Proposition 6.1]{OV}), we can put all of the above together and conclude that it suffices to show
\begin{equation}\label{inverse_norm_sum}
S(\Lambda^0,s,R) = \frac{w_K}{R_KN(\frak{a})^s}\zeta_K^{[\frak{a}]^{-1}}(s)c_n\log(R)^{n-1} + O(\log(R)^{n-2})
\end{equation}
from which the theorem will follow immediately.

Let us write the dominant error term in the approximation (\ref{term_approx}) for $b_{k,R}^{\frak{a}}$ as $c_k^\frak{a}\log(R^n/k)^{n-2}$, for some constant $c_k^\frak{a}$ which may depend on $n$, $k$, and $\frak{a}$ but not on $R$.  In that case we can write, using Theorem \ref{bkR_estimate},
\begin{align}\label{simplify_ins}
S(\Lambda^0,s,R) &= \sum_{k = 1}^{R^n} \frac{b_{k,R}^\frak{a}}{k^s} \\
&= \frac{w_K}{R_K(n-1)!}\left(\sum_{k = 1}^{R^n}\frac{a_{k}^{[1],\frak{a}} }{k^s}\log(R^n/k)^{n-1} + \sum_{k = 1}^{R^n}\frac{c_k^\frak{a}}{k^s}\log(R^n/k)^{n-2}\right) + \text{smaller terms}
\end{align}
Let us begin to analyze this expression by concentrating on the first summation inside the parentheses.  First, recall that the norm of any principal ideal contained in $\frak{a}$ must have norm a multiple of $N = N(\frak{a})$.  We have now, by reindexing and using Lemma \ref{prinideal_coeffs},
\begin{align} 
\sum_{k = 1}^{R^n}\frac{a_{k}^{[1],\frak{a} }}{k^s}\log(R^n/k)^{n-1} 
&= \sum_{k = 1}^{\left\lfloor R^n/N\right\rfloor}\frac{a_{kN}^{[1],\frak{a}}}{(kN)^s}\log(R^n/kN)^{n-1} \\
&= \frac{1}{N^s}\sum_{k = 1}^{\left\lfloor R^n/N\right\rfloor}\frac{a_k^{[\frak{a}]^{-1}} }{k^s}(\log(R^n)-\log(kN))^{n-1} \\
&= \frac{1}{N^s}\sum_{k = 1}^{\left\lfloor R^n/N\right\rfloor}\frac{a_k^{[\frak{a}]^{-1}} }{k^s}\sum_{m = 0}^{n-1}(-1)^m\binom{n-1}{m}\log(R^n)^{n-1-m}\log(kN)^m \\
&= \frac{1}{N^s}\sum_{m = 0}^{n-1}\left[\binom{n-1}{m}\log(R^n)^{n-1-m}\sum_{k = 1}^{\left\lfloor R^n/N\right\rfloor}(-1)^m\frac{a_k^{[\frak{a}]^{-1}} \log(kN)^m}{k^s}\right]
\end{align}

When $m = 0$, then corresponding summand in the above is
\begin{align}
\frac{1}{N^s}\log(R^n)^{n-1}\sum_{k = 1}^{\left\lfloor R^n/N\right\rfloor}\frac{a_k^{[\frak{a}]^{-1}}}{k^s} &= \frac{1}{N^s}\log(R^n)^{n-1}\left(\zeta_K^{[\frak{a}]^{-1}}(s) - \sum_{k = 1}\frac{a_k^{[\frak{a}]^{-1}}}{k^s}\right) \\
&= \frac{1}{N^s}\log(R^n)^{n-1}\zeta_K^{[\frak{a}]^{-1}}(s) + O(1)
\end{align}
where we have used Lemma \ref{ekaarvio} to estimate the tail of the ideal class zeta function. When $m>0$, we can use Lemma \ref{ekaarvio} again to establish the easy bounds
\begin{align}
\sum_{k = 1}^{\left\lfloor R^n/N\right\rfloor}(-1)^m\frac{a_k^{[\frak{a}]^{-1}} \log(kN)^m}{k^s} &= \sum_{k=1}^{\infty}(-1)^m\frac{a_k^{[\frak{a}]^{-1}}\log(kN)^m}{k^s} - \sum_{k = \left\lfloor R^n/N\right\rfloor + 1}^{\infty}(-1)^m\frac{a_k^{[\frak{a}]^{-1}}\log(kN)^m}{k^s} \\
&\leq \sum_{k = 1}^{\infty} (-1)^m\frac{a_k^{[\frak{a}]^{-1}}\log(kN)^m}{k^s} + (m+1)\log(N)^m\sum_{k = \left\lfloor R^n/N\right\rfloor + 1}^{\infty}\frac{a_k^{[\frak{a}]^{-1}}\log(k)^m}{k^s} \\
&\leq \max_{m = 0,\ldots,n} \left\{(m+1)\log(N)^m|\zeta_K^{[\frak{a}]^{-1},(m)}(s)|\right\}
\end{align}
where the second-to-last inequality comes from writing out $\log(kN)^m = (\log(k) + \log(N))^m$ in a binomial expansion.  Substituting these estimates back into the sum of interest, we arrive at 
\begin{equation}
\sum_{k = 1}^{R^n}\frac{a_k^{[1],\frak{a}}}{k^s}\log(R^n/k)^{n-1} = \frac{1}{N(\frak{a})^s}\zeta_K^{[\frak{a}]^{-1}}(s)n^{n-1}\log(R)^{n-1} + O(\log(R)^{n-2})
\end{equation}

We now extract the error term and rewrite it in a similar manner.  Since the regions $\mathcal{S}_k$ in the proof of Theorem \ref{bkR_estimate} are all scaled version of $\mathcal{S}_1$, and the lattices whose points we are counting are all translated versions of $\Lambda_{\log}$, it follows from \cite[Chapter VI \S2, Theorem 2]{lang} that we can find a constant $c$ independent of $k$ such that $c_k^\frak{a}\leq ca_k^{1,\frak{a}}$ for all $k$.  We get
\begin{equation}
\sum_{k = 1}^{R^n}\frac{c_k^\frak{a}}{k^s}\log(R^n/k)^{n-2} \leq c\sum_{k = 1}^{R^n}\frac{a_k^{1,\frak{a}}}{k^s}\log(R^n/k)^{n-2} = O(\log(R)^{n-2})
\end{equation}
as claimed.  Again, the last equality follows from writing out the binomial expansion of $\log(R^n/k)$ as above, and using Lemma \ref{ekaarvio}, which shows that the error introduced by including the tail of the zeta function is minuscule when compared to $\log(R)^{n-2}$.  Plugging all of the above back into (\ref{simplify_ins}) completes the proof of the theorem.
\end{IEEEproof}

We can use the part of the coefficient of $\log(R)^{n-1}$ in Theorem \ref{main_theorem} which depends on the specific ideal lattice to define the following invariant of $\Lambda = (\frak{a},q_\alpha)$:
\begin{equation}
\boxed{\sigma(K,[\frak{a}],s) = \frac{w_K|D_K|^{s/2}}{R_K}\zeta_K^{[\frak{a}]^{-1}}(s)}
\end{equation}
which depends only on $K$ and the ideal class $[\frak{a}]$, which are in turn enough to determine the growth of the inverse norm sum.  To compare the inverse norm sums of two normalized ideal lattices of the same dimension, one must now only look at the coefficient $\sigma(K,[\frak{a}],s)$.  Note that there is no dependence on the twisting element $\alpha$.

\begin{example} \emph{Real Quadratic Fields}.  
Let us consider the fields $\Q(\sqrt{d})$ with $d>0$ and ideal lattices of the form $\Lambda = (\mathcal{O}_K,q_\alpha)$ as in \cite{ducoat_oggier}.  One can predict the value of $S(\Lambda,s,R)$ from the formula
\begin{equation}
S_K^{(\mathcal{O}_K,q_\alpha)}(s,R) \approx 2\sigma(K,[1],s)\log(R)
\end{equation}
The corresponding ranking of fields for $s = 3$ is given in Table \ref{scaled_real_quad_table}.  The fields were taken from Table I of \cite{ducoat_oggier}, wherein inverse norm sums for normalized lattices of the form $(\mathcal{O}_K,q_\alpha)$ were computed for $R = 100$.  
\begin{table}[h!] 
\centering
\caption{Real quadratic fields $\Q(\sqrt{d})$ for $d\leq 100$, ordered according to $\sigma(K,[1],3)$} \label{scaled_real_quad_table}
\begin{tabular}{|c|c|c|c|c|c|c|c|c|}
\hline
$d$ & $h_K$ & $R_K$ & $D_K$ & $\zeta_K^{[1]}(3)$ & $\sigma(K,[1],3)$ & 
\begin{tabular}{cc}
\text{Predicted} \\ $S_K^{(\mathcal{O}_K,q_\alpha)}(3,100)$ 
\end{tabular}
& 
\begin{tabular}{cc}
\text{Actual} \\ $S_K^{(\mathcal{O}_K,q_\alpha)}(3,100)$
\end{tabular}
&
\text{error (\%)}
\\
\hline
 5 & 1 & 0.4812 & 5 & 1.0275 & 47.7475 & 439.8 & 458.1 & 4.0 \\
 \hline
 2 & 1 & 0.8814 & 8 & 1.1520 & 59.1518 & 544.8 & 611.4 & 10.9 \\
 \hline
 13 & 1 & 1.1948 & 13 & 1.0969 & 86.0647 & 792.7 & 821.7 & 3.5 \\
 \hline
 17 & 1 & 2.0947 & 17 & 1.3100 & 87.6679 & 807.5 & 1049.8 & 23.1 \\
 \hline
 41 & 1 & 4.1591 & 41 & 1.3296 & 167.8478 & 1545.9 & 1535.7 & 0.7 \\
 \hline
 29 & 1 & 1.6472 & 29 & 1.0410 & 197.3910 & 1818.0 & 1945.0 & 6.5 \\
 \hline
 37 & 1 & 2.4918 & 37 & 1.1038 & 199.3926 & 1836.5 & 1985.6 & 7.5 \\
 \hline
 10 & 2 & 1.8184 & 40 & 1.0315 & 287.0103 & 2643.5 & 3121.8 & 15.3\\
\hline
\end{tabular}
\end{table}
Note that the invariant $\sigma(K,[1],s)$ suffices to order the fields according to their inverse norm sums (although the correct ordering between $d = 29$ and $d = 37$ is likely an accident, since the difference between the actual inverse norm sums is so small compared to the error of our approximation).  Lastly, as is noted in \cite{ducoat_oggier}, evaluating inverse norm sums is computationally burdensome and dependent on $R$, whereas $\sigma(K,[1],s)$ is simple to calculate provided one knows the basic invariants of $K$.
\end{example}

\begin{example} \emph{Real Quartic Fields}.
We repeat the above experiment for the real quartic fields $K_1,\ldots,K_6$ given in Table III of \cite{ducoat_oggier}, whose minimal polynomials are defined therein.  The fields are ranked below in Table \ref{scaled_real_quartic_table} according to $\sigma(K,[1],3)$.  
\begin{table}[h!] 
\centering
\caption{Real quartic fields from Table III of \cite{ducoat_oggier}, ordered according to $\sigma(K,[1],3)$} \label{scaled_real_quartic_table}
\begin{tabular}{|c|c|c|c|c|c|}
\hline
Field & $h_K$ & $R_K$ & $D_K$ & $\zeta_K^{[1]}(3)$ & $\sigma(K,[1],3)$ \\
\hline
 $K_1$ & 1 & 0.8251 & 725 & 1.0023 & 47429 \\
 \hline
 $K_2$ & 1 & 1.1655 & 1125 & 1.0100 & 65404 \\
 \hline
 $K_3$ & 1 & 1.0190 & 1600 & 1.0190 & 84556 \\
 \hline
 $K_6$ & 1 & 1.1440 & 2048 & 1.1440 & 86847 \\
 \hline
 $K_4$ & 1 & 1.9184 & 1957 & 1.0422 & 94066 \\
 \hline
 $K_5$ & 1 & 1.8528 & 2000 & 1.0422 & 98941 \\
\hline
\end{tabular}
\end{table}
Upon comparing the values of the corresponding inverse norm sums for $R = 5$ as tabulated in Table III of \cite{ducoat_oggier}, we see that the ranking provided by the invariant $\sigma(K,[1],3)$ is exactly the same as that given by the inverse norm sum.  Thus $\sigma(K,[1],3)$ suffices to predict the relative behavior of the inverse norm sums of these fields.  We should also remark that one could use Theorem \ref{main_theorem} to predict the actual value of $S(\Lambda,s,R)$.  However, the error in doing so appears quite large, which we attribute to the small value of $R$ relative to the dimension and the slow growth of the function $\log(R)^{n-1}$.
\end{example}

The above tables and examples do not give the whole picture for real quadratic and quartic fields, since we have only considered principal ideal classes.  If one were to consider ideal lattices $(\frak{a},q_\alpha)$ such that $[\frak{a}]\neq [1]$, then the zeta values $\zeta_K^{[\frak{a}]^{-1}}(s)$ will be remarkably different, likely changing the outcome of such an experiment.  We use the next two examples to see how $\zeta_K^{[\frak{a}]^{-1}}(s)$ behaves with respect to varying $[\frak{a}]$.

\begin{example}
Let us consider the number field $K = \Q(\sqrt{229})$ with ring of integers $\mathcal{O}_K = \Z[\omega]$,  $\omega = (1+\sqrt{229})/2$ and class number $h_K = 3$.  Let $\sigma$ be the non-trivial element of the Galois group $\text{Gal}(K/\Q)$.  The class group $C_K$ can be described by
\begin{equation}
C_K = \{[\frak{a}_1]=[1],[\frak{a}_2],[\frak{a}_3]\},\	\text{where}\ \frak{a}_1 = (1),\ \frak{a}_2 = (3,\omega),\ \text{and}\ \frak{a}_3 = (3,\sigma(\omega))
\end{equation}
We consider three ideal lattices $\Lambda_i = (\frak{c}_i,q_{\alpha_i})$, where
\begin{equation}
\frak{c}_1 = (-2+\sqrt{229}),\ \frak{c}_2 = \left(225,(173+\sqrt{229})/2\right),\ \text{and}\ \frak{c}_3 = \left(75,(69+3\sqrt{229})/2\right)
\end{equation}
Let us compare the growth of the inverse norm sums corresponding to $\Lambda_i$.  The ideals $\frak{c}_i$ were chosen because they all satisfy $N(\frak{c}_i) = 225$, and hence their canonical embeddings (taking, for example, $\alpha = 1$) all give lattices of the same volume.  However they all represent different ideal classes.  Indeed, we have $[\frak{c}_i] = [\frak{a}_i]$ for $i = 1,2,3$ in the ideal class group.  

The only term that differentiates the coefficients $\sigma(K,[\frak{a}_i],s)$, and thus the growth of the corresponding inverse norm sums, is the value of the zeta function $\zeta_K^{[\frak{a}_i]^{-1}}(s)$.  These values for $s = 2$ and $s = 3$ are tabulated in Table \ref{zeta_value_table} below.
\begin{table}[h!] 
\centering
\caption{Values of $\sigma(K,[\frak{a}_i],s)$ for the field $K = \Q(\sqrt{229})$} \label{zeta_value_table}
\begin{tabular}{|c|c|c|c|c|}
\hline
 Ideal class $[\frak{a}]$ & $\zeta_K^{[\frak{a}]}(2)$ & $\sigma(K,[\frak{a}],2)$ & $\zeta_K^{[\frak{a}]}(3)$ & $\sigma(K,[\frak{a}],3)$ \\
\hline
$[\frak{a}_1] = [1]$ & 1.1056 & 186.6807 & 1.0182 & 171.9232 \\
\hline
$[\frak{a}_2]$ & 0.2061 & 34.8000 & 0.0488 & 8.2399 \\
\hline
$[\frak{a}_3]$ & 0.2061 & 34.8000 & 0.0488 & 8.2399 \\
\hline
\end{tabular}
\end{table}
From these results we can see that ideal lattices built over the non-principal ideals $\frak{c}_2$ and $\frak{c}_3$ will have much smaller inverse norm sums.  We are not claiming that the resulting lattices $\Lambda_i = (\frak{a}_i,q_{\alpha_i})$ are optimal in any sense for the wiretap channel, only presenting evidence that everything else equal, one may prefer lattices coming from non-principal ideals due to the much smaller zeta values.
\end{example}

Notice that the values $\zeta_K^{[\frak{c}_i]}(s)$ are the same for $i = 2,3$ in the above table, which can be explained as follows.  For any Galois extension $K/\Q$, the group $\text{Gal}(K/\Q)$ acts on $C_K$ in an obvious way, namely by $\sigma([\frak{a}]) = [\sigma(\frak{a})]$.  Since Galois action preserves norms of ideals, one can show easily that $\zeta_K^{[\frak{a}]}(s) = \zeta_K^{[\sigma(\frak{a})]}(s)$ for all $\sigma\in \text{Gal}(K/\Q)$.  In the above example we have $[\frak{c}_3] = \sigma([\frak{c}_2])$.  Knowing that two ideal classes are Galois conjugate reduces computational tasks, since one only needs to compute zeta values for one representative in each orbit of $\text{Gal}(K/\Q)$ on $C_K$.

\begin{example}
Let $K = \Q[X]/(f(X))$ where $f(X) = X^4 - 200X^2 + 324$ and let $\omega$ be a root of $f$.  The class group $C_K$ is cyclic of order $6$, with representatives
\begin{align}
\frak{a}_0 &= (1) \\
\frak{a}_1 &= (10, 7\omega^3/72 + 3\omega^2/4 - 691\omega/36 - 72) \\
\frak{a}_2 &= (50, -23\omega^3/72 - \omega^2/2 + 2849\omega/36 + 51/2) \\
\frak{a}_3 &= (2, \omega^3/36 + \omega^2/4 - 50\omega/9 - 51/2) \\
\frak{a}_4 &= (5, -\omega^3/72 + \omega^2/4 + 73\omega/36 - 24) \\
\frak{a}_5 &= (50, 3\omega^3/8 - \omega^2/2 - 289\omega/4 + 101/2)
\end{align}
The group $C_K$ is generated by $[\frak{a}_1]$ and we have $[\frak{a}_i] = [\frak{a}_1]^i$ for all $i = 0,\ldots,5$.  Thus $[\frak{a}_5]$ also generates $C_K$, $[\frak{a}_2]$ and $[\frak{a}_4]$ have order $3$, and $[\frak{a}_3]$ is the lone element of order $2$.  In fact, $[\frak{a}_1]$ and $[\frak{a}_5]$ are Galois conjugate, and so are $[\frak{a}_2]$ and $[\frak{a}_4]$.  The values of the corresponding ideal class zeta functions are tabulated in Table \ref{zeta_value_table3}.
\begin{table}[h!] 
\centering
\caption{Values of $\sigma(K,[\frak{a}],s)$ for the field $K = \Q[X]/(X^4-200X^2+324)$} \label{zeta_value_table3}
\begin{tabular}{|c|c|c|c|c|}
\hline
Ideal class $[\frak{a}]$ & $\zeta_K^{[\frak{a}]}(2)$ & $\sigma(K,[\frak{a}],2)$ & $\zeta_K^{[\frak{a}]}(3)$ & $\sigma(K,[\frak{a}],3)$ \\
\hline
$[\frak{a}_0] = [1]$ & 1.2358 & 4.60$\times 10^6$ & 1.0492 & 4.45$\times10^9$ \\
\hline
$[\frak{a}_1]$ & 0.0595 & 2.21$\times 10^4$ & 0.0044 & 1.58$\times 10^7$ \\
\hline
$[\frak{a}_2]$ & 0.1126 & 4.19$\times 10^4$ & 0.0172 & 6.19$\times 10^7$ \\
\hline
$[\frak{a}_3]$ & 0.6059 & 2.25$\times 10^5$ & 0.2610 & 9.40$\times 10^8$ \\
\hline
\end{tabular}
\end{table}
Note that the value of the ideal class zeta function is inversely related to the order of the corresponding ideal class in $C_K$.  We believe this is evidence of a general phenomenon, but leave further consideration along these lines for future work.  A more thorough analysis will involve explicit calculation of the actual inverse norm sums, which we also save for future work.
\end{example}

\begin{example}
Consider the field $K = \Q[X]/(X^4-X^3-3X^3+X+1)$ and the ideal lattice $\Lambda = (\mathcal{O}_K,q_1)$ corresponding to the full ring of integers, with $R = 10$.  Using the notation of the examples of Section \ref{geometric}, our main theorem says that up to a multiplicative constant $c$, we can approximate the PEP (cf.\ \ref{geometric}) by
\begin{equation}
\frac{1}{\gamma^n}\sum_{k = 1}^{R^4}\frac{b_{k,R}}{k^2} \approx \frac{1}{\gamma^n}\sum_{k = 1}^{R^4}\frac{n_{k,R}}{k^2}
\end{equation}
where $\gamma$ is the average SNR.  In Fig.\ \ref{wiretap_snr} we plot the standard PEP curves, ignoring the constant $c$ which is the same for both sums, and letting $\gamma$ take values over an SNR range.  The figure shows that there is no penalty in using the estimates $n_{k,R}$ in place of the exact values $b_{k,R}$ when computing the PEP.
\begin{figure}[h!]
\begin{center}
\includegraphics[width=.75\textwidth]{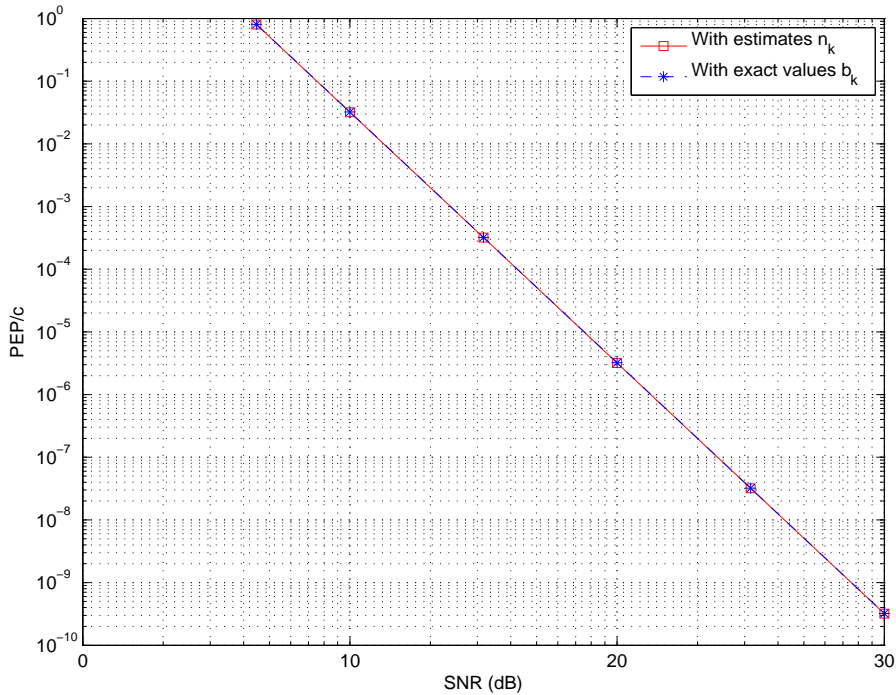}
\end{center}
\caption{PEP/$c$ as a function of SNR using the approximation given by Theorem \ref{main_theorem}, i.e.\ the estimates $n_{k,R}$ and 2) the exact inverse norm sum, i.e.\ the values $b_{k,R}$.  The field is $K = \Q[X]/(X^4-X^3-3X^3+X+1)$ with ideal lattice $\frak{a}=\mathcal{O}_K$ and $R = 10$.}
\label{wiretap_snr}
\end{figure}
\end{example}

\section{Conclusions and future work}\label{conclusions}

We have considered lattice codes from ideal lattices constructed over totally real algebraic number fields.  Our main theorem, Theorem \ref{main_theorem}, provides an estimate of the corresponding inverse norm sum when we normalize the lattice to have unit volume.  This allows us to determine the exact number theoretic invariants on which the inverse norm sum depends.  In particular, we have showed a heavy dependence on the values of ideal class Dedekind zeta functions, and that in some cases considering non-principal ideals may be beneficial due to their small zeta values.  Along the way, we derived an estimate for the number of constellation points with certain algebraic norm in a given ideal, the accuracy of which was demonstrated through practical examples.

Future work will consist of generalizing the results to complex lattices and multiple-input multiple-output (MIMO) channels.   For a CM-field $K$ with $K = K'L$, $K'$ totally real, and $L$ quadratic imaginary, one can study the relative embedding $K\hookrightarrow \C^n$ which fixes a given embedding of $L$.  The corresponding inverse norm sum can likely be similarly analyzed as in this paper.  One promising approach is offered by division algebras, along the same lines as in  \cite{roopefransujournal,roopelaura_ISIT}, and one could potentially generalize the theorems therein using methods similar to ours. In addition, for the wiretap channel we have only concentrated on the design of the eavesdropper's lattice, while in truth we must simultaneously design the legitimate user's lattice as well.  Lastly, a deeper numerical analysis of our results and potentially creating good lattice codes from non-principal ideals will require computing the corresponding inverse norm sums explicitly and finally simulating the codes.

\section{Acknowledgments} The authors would like to thank Prof. Fr\'ed\'erique Oggier, Prof. Jean-Claude Belfiore, and Dr. Roope Vehkalahti for useful discussions, as well as the anonymous reviewers who's comments greatly improved the quality and exposition of this paper.


\begin{thebibliography}{10}

\bibitem{ICUMT11}
C.~Hollanti and E.~Viterbo,
\newblock ``Analysis on wiretap lattice codes and probability bounds from
  {Dedekind} zeta functions'',
\newblock in {\em 3rd International Congrass on Ultra Modern Telecommunications
  and Control Systems and Workshops (ICUMT)}, 2011.

\bibitem{OV}
F.~Oggier and E.~Viterbo,
\newblock {\em Algebraic number theory and code design for Rayleigh fading
  channels}, vol. 1, issue 3 of {\em Foundations and Trends in Communications
  and Information Theory},
\newblock Now Publishers Inc., Hanover, MA, USA, December 2004.

\bibitem{Hell_Gaussianwire}
S.~Leung-Yan-Cheong and M.~Hellman,
\newblock ``The {Gaussian} wire-tap channel'',
\newblock {\em IEEE Transactions on Information Theory}, vol. 24, no. 4, pp.
  451--456, July 1978.

\bibitem{belfisoleoggis}
F.~Oggier, P.~Sol{\'e}, and J.-C. Belfiore,
\newblock ``Lattice codes for the wiretap {Gaussian} channel: {C}onstruction
  and analysis'',
\newblock 2013,
\newblock arxiv.1103.4086.

\bibitem{belfisole}
J.-C. Belfiore and P.~Sol{\'e},
\newblock ``Unimodular lattices for the {Gaussian} wiretap channel'',
\newblock in {\em IEEE Information Theory Workshop (ITW)}, 2010.

\bibitem{belfioggiswire}
J.-C. Belfiore and F.~E. Oggier,
\newblock ``Secrecy gain: {A} wiretap lattice code design'',
\newblock in {\em International Symposium on Information Theory and its
  Applications (ISITA)}, 2010.

\bibitem{oggis_new}
J.-C. Belfiore and F.~Oggier,
\newblock ``An error probability approach to {MIMO} wiretap channels'',
\newblock {\em IEEE Transactions on Communications}, vol. 61, no. 8, pp.
  3396--3403, June 2013.

\bibitem{BO_wiretap}
J.-C. Belfiore and F.~Oggier,
\newblock ``Lattice code design for the rayleigh fading wiretap channel'',
\newblock in {\em IEEE International Conference on Communications (ICC)}, 2011.

\bibitem{ITW_camiame}
A.-M. Ernvall-Hyt\"onen and C.~Hollanti,
\newblock ``On the eavesdropper's correct decision in {Gaussian} and fading
  wiretap channels using lattice codes'',
\newblock in {\em IEEE Information Theory Workshop (ITW)}, 2011.

\bibitem{ISIT11-roopefransu}
R.~Vehkalahti and H.-F.~(F.) Lu,
\newblock ``An algebraic look into {MAC-DMT} of lattice space-time codes'',
\newblock in {\em IEEE International Symposium on Information Theory (ISIT)},
  2011.

\bibitem{ITW11-roopefransu}
R.~Vehkalahti and H.-F.~(F.) Lu,
\newblock ``Diversity-multiplexing gain tradeoff: a tool in algebra?'',
\newblock in {\em IEEE Information Theory Workshop (ITW)}, 2011.

\bibitem{roopelaura_ISIT}
R.~Vehkalahti and L.~Luzzi,
\newblock ``Connecting {DMT} of division algebra space-time codes and point
  counting in {Lie} groups'',
\newblock in {\em IEEE International Symposium on Information Theory (ISIT)},
  2012.

\bibitem{roopefransujournal}
R.~Vehkalahti, H.-F.~(F.) Lu, and L.~Luzzi,
\newblock ``Inverse determinant sums and connections between fading channel
  information theory and algebra'',
\newblock {\em {IEEE} Trans. Inf. Theory}, vol. 59, no. 9, pp. 6060--6082,
  September 2011.

\bibitem{ducoat_oggier}
J.~Ducoat and F.~Oggier,
\newblock ``An analysis of small dimensional fading wiretap lattice codes'',
\newblock in {\em IEEE International Symposium on Information Theory (ISIT)},
  2014.

\bibitem{ong_oggier}
S.~Ong and F.~Oggier,
\newblock ``Wiretap lattice codes from number fields with no small norm
  elements'',
\newblock {\em Designs, Codes, and Cryptography}, vol. 73, no. 2, pp. 425--440,
  November 2014.

\bibitem{EV}
G.~R. Everest,
\newblock ``On the solution of the norm-form equation'',
\newblock {\em Amer. J. Math.}, vol. 114, no. 3, pp. 667--682, 1992.

\bibitem{EVLO}
G.~Everest and J.H. Loxton,
\newblock ``Counting algebraic units with bounded height'',
\newblock {\em J. Number Theory}, vol. 44, pp. 222--227, 1993.

\bibitem{lang}
S.~Lang,
\newblock {\em Algebraic number theory},
\newblock Springer-Verlag New York Inc., 1986.

\bibitem{sage}
``{Sage} open source mathematics software system'',
\newblock http://www.sagemath.org/.

\bibitem{Wyner}
A.~Wyner,
\newblock ``The wire-tap channel'',
\newblock {\em Bell. Syst. Tech. Journal}, vol. 54, 1975.

\end{thebibliography}

\newpage
\section*{Appendix}
We devote the appendix to proving Lemma \ref{ekaarvio}:

\emph{Lemma 2:} Suppose that $R \geq 3$, let $[\frak{a}]$ be an ideal class in $K$, and let $N$ be a constant.  We have
\begin{equation}
\sum_{k= (R^n +1)/N}^{\infty}\frac{a^{[\frak{a}]}_k(\log (kN))^m}{k^s}\leq \begin{cases}c R^{-n}(\log (R^n))^m,\quad&\textrm{when $s=2$}\\ cR^{-2n}(\log (R^n))^m,\quad & \textrm{when $s=3$}\end{cases}
\end{equation}
where $c$ is a constant depending on the field $K$ and the ideal $\mathfrak{a}$, but not on $R$.

\begin{IEEEproof} Throughout the proof, we may assume that $R$ is large, because if we are able to prove the existence of such a constant $c$ for large enough $R$, then we can find a constant $c$ suitable for all values of $R$ by treating the small values by comparing the values of the sum on the left hand side of the inequality, and the expression on the right hand side of the inequality.

By \cite[Chapter VI, \S3, Theorem 3]{lang}, we have
\begin{equation}\label{sum_coeffs}
\sum_{k\leq t}a^{[\frak{a}]}_k =\kappa t+O(t^{1-1/n})
\end{equation}
for some constant $\kappa$ depending on $K$ and $\mathfrak{a}$.  For simplicity, denote $T:=\frac{R^n+1}{N}$. Let us split the interval
\begin{equation}
[T,\infty)=\cup_{h=0}^{\infty}[2^hT,2^{h+1}T).
\end{equation}

Now the aim is to show that we can use geometric sums to estimate the sum in question, and in particular, that we can form the geometric sums in such a way that every interval in the dyadic splitting yields one term.

We have
\[
\sum_{2^hT\leq k<2^{h+1}Tt}a^{[\frak{a}]}_k=\kappa 2^{h}T+O\left(\left(2^{h+1}T\right)^{1-1/n}\right).
\]
Let us now consider the function
\[
f(x)=\frac{(\log(xN))^m}{x^s}.
\]
Now
\[
f'(x)=m\frac{(\log(xN))^{m-1}}{x^s+1}-s\frac{(\log(xN))^m}{x^{s+1}}=\frac{(\log(xN))^{m-1}}{x^{s+1}}(m-s\log(xc))=0,
\]
when $m=s\log(xN)$, that is, when $x=\frac{e^{m/s}}{N}$, and hence, the function is decreasing the interval we are considering.

We may thus estimate:
\begin{align*}
\sum_{2^hT\leq k<2^{h+1}T}\frac{a^{[\frak{a}]}(\log(kN))^m}{k^{s}}\leq \frac{(\log(2^hTN))^m}{(2^hT)^s}\sum_{2^hT\leq k<2^{h+1}T}a^{[\frak{a}]}_k\\=\frac{(\log(2^hTN))^m}{(2^hT)^s}\left(\kappa 2^hT+O\left(\left(2^{h+1}T\right)^{1-1/n}\right)\right)
\end{align*}
Finally, we need to sum over the values of $h$. Let us start from the error term:
\[
\sum_{h\geq 0}\frac{(\log(2^hTN))^m}{(2^hT)^s}\left(2^{h+1}T\right)^{1-1/n}\leq \sum_{h\geq 0}\frac{(\log(TN))^m}{T^s}\left(2^{h+1}T\right)^{1-1/n} =O\left(\frac{(\log(TN))^m}{T^{s-1+1/n}}\right)
\]
We may now turn to the main term. We want to now show that the main terms can be majored by a geometric progression. To do so, let us consider the ratio between two consecutive main terms. We have
\begin{align*}
\frac{(\log(2^{h+1}TN))^m(2^{h+1}T)^{1-s}}{(\log(2^hTN))^m(2^hT)^{1-s}}=2^{1-s}\left(\frac{\log(2^{h+1}TN)}{\log(2^hTN)}\right)^m=2^{1-s}\left(\frac{\log 2}{\log(2^hTN)}+1\right)^m.
\end{align*}
Since T is large,
\[
\frac{\log 2}{\log(2^hTN)}<\frac{1}{2m},
\]
and hence,
\[
\left(\frac{\log 2}{\log(2^hTN)}+1\right)^m<\left(\frac{1}{2m}+1\right)^m<e^{1/2}<1.7.
\]
Thus,
\[
\frac{(\log(2^{h+1}TN))^m(2^{h+1}T)^{1-s}}{(\log(2^hTN))^m(2^hT)^{1-s}}<2^{1-s}\cdot 1.7\leq \frac{1.7}{2}<1.
\]
We may thus estimate the sum as a geometric progression:
\begin{align*}
\sum_{2^hT\leq k}\frac{a^{[\frak{a}]}(\log(kN))^m}{k^{s}}=\sum_{h\geq 0}\sum_{2^hT\leq k<2^{h+1}T}\frac{a^{[\frak{a}]}(\log(kN))^m}{k^{s}}\\ \leq \sum_{h\geq 0} \left(\frac{1.7}{2}\right)^h \kappa T\frac{(\log(TN))^m}{T^s}=O\left(T\frac{(\log(TN))^m}{T^s}\right),
\end{align*}
which completes the proof.
\end{IEEEproof}
\end{document}